\documentclass{aa}
\usepackage{graphicx}
\usepackage{txfonts}
\usepackage{color}
\usepackage{natbib}
\usepackage{ulem}
\bibpunct{(}{)}{;}{a}{}{,} % to follow the A&A style
\usepackage{hyperref}
\begin{document}

    \titlerunning{Spectropolarimetry of Type II SNe (II)}  
    \authorrunning{T. Nagao et al.}

   \title{Spectropolarimetry of Type~II supernovae}
   %Diverse explosion geometries of Type II supernovae implied by their diverse continuum-polarization properties}
   %Diversity in the continuum polarization of Type II supernovae: Implication for their diverse explosion geometries}

   \subtitle{(II) Intrinsic supernova polarization and its relations with the photometric/spectroscopic properties}

   \author{
            T.~Nagao,\inst{1,2,3}\fnmsep\thanks{takashi.nagao@utu.fi} % 0000-0002-3933-7861
            F.~Patat,\inst{4} % 0000-0002-0537-3573
            A.~Cikota,\inst{5} % 0000-0001-7101-9831
            D.~Baade,\inst{4} % 0000-0003-1637-9679
            S.~Mattila,\inst{1,6} % 0000-0001-7497-2994
            R.~Kotak,\inst{1} % 0000-0001-5455-3653
            H.~Kuncarayakti,\inst{1,7} % 0000-0002-1132-1366
            M.~Bulla,\inst{8,9,10} % 0000-0002-8255-5127
            \and
            B.~Ayala,\inst{11}
          }

   \institute{
            Department of Physics and Astronomy, University of Turku, FI-20014 Turku, Finland
            \and
            Aalto University Mets\"ahovi Radio Observatory, Mets\"ahovintie 114, 02540 Kylm\"al\"a, Finland
            \and
            Aalto University Department of Electronics and Nanoengineering, P.O. BOX 15500, FI-00076 AALTO, Finland
            \and
            European Southern Observatory, Karl-Schwarzschild-Str. 2, 85748 Garching b. M\"{u}nchen, Germany
            \and
            Gemini Observatory/NSF’s NOIRLab, Casilla 603, La Serena, Chile
            \and
            School of Sciences, European University Cyprus, Diogenes street, Engomi, 1516 Nicosia, Cyprus
            \and
            Finnish Centre for Astronomy with ESO (FINCA), FI-20014 University of Turku, Finland
            \and
            Department of Physics and Earth Science, University of Ferrara, via Saragat 1, I-44122 Ferrara, Italy
            \and
            INFN, Sezione di Ferrara, via Saragat 1, I-44122 Ferrara, Italy
            \and 
            INAF, Osservatorio Astronomico d’Abruzzo, via Mentore Maggini snc, 64100 Teramo, Italy
            \and
            Departamento de Ciencias Fisicas, Universidad Andres Bello, Avda. Republica 252, Santiago, Chile
             }

   \date{Received April ??, 2023; accepted ?? ??, 2023}

% \abstract{}{}{}{}{} 
% 5 {} token are mandatory
 
  \abstract
  {The explosion processes of supernovae (SNe) are imprinted in their explosion geometries. The recent discovery of several highly aspherical core-collapse SNe is significant, and regarded as the key to understanding the explosion mechanism of core-collapse SNe. Here, we study the intrinsic polarization of 15 hydrogen-rich core-collapse SNe and explore the relation with the photometric and spectroscopic properties. 
  Our sample shows diverse properties of the continuum polarization. The polarization of most SNe has a low degree at early phases but shows a sudden rise to $\sim 1$ \% degree at certain points during the photospheric phase as well as a slow decline during the tail phase, with a constant polarization angle. The variation in the timing of peak polarisation values implies diversity in the explosion geometry: some SNe have aspherical structures only in their helium cores, while in other SNe these reach out to a significant part of the outer hydrogen envelope with a common axis from the helium core to the hydrogen envelope.
  Other SNe show high polarization from early phases and a change of the polarization angle around the middle of the photospheric phase. This implies that the ejecta are significantly aspherical to the outermost layer and have multi-directional aspherical structures. %These SNe might originate from a combination of a common aspherical SN explosion and an aspherical circumstellar-material (CSM) interaction or from a multi-directional SN explosion. 
  Exceptionally, the Type~IIL SN~2017ahn shows low polarization at both the photospheric and tail phases. %Since we have only one such example, we cannot conclude whether this SN comes from a completely different type of explosion or whether it is simply the effect of a different viewing angle for a similarly aspherical explosion.
  Our results show that the timing of the polarization rise in Type~IIP SNe is likely correlated with their brightness, velocity and the amount of radioactive Ni produced: brighter SNe with faster ejecta velocity and a larger $^{56}$Ni mass have more extended-aspherical explosion geometries. In particular, there is a clear correlation between the timing of the polarization rise and the explosion energy, that is, the explosion asphericity is proportional to the explosion energy. This implies that the development of a global aspherical structure, e.g., a jet, might be the key to realising an energetic SN in the mechanism of SN explosions.
  }
  
%   {TBD;
%   aaa}
%   {TBD;
%   {TBD;


   \maketitle

%
%-------------------------------------------------------------------

\section{Introduction} \label{sec:intro}

Core-collapse supernovae (CCSNe) are catastrophic explosions of massive stars at their deaths. This phenomenon is closely related to the evolution of galaxies through the processes such as the chemical enrichment of galaxies and the induction of star formation. Despite great efforts, the explosion mechanism has not fully been understood yet. The most promising picture is the neutrino-driven explosion, in which a star blows up due to neutrino heating from a proto-neutron star \citep[e.g.,][]{Janka2007,Janka2012}. Even though some recent multi-dimensional simulations have been successful in launching the explosion, they are unable to reproduce some of the basic observed properties, e.g., the $10^{51}$ erg of explosion energy \citep[e.g.,][]{Buras2006, Marek2009, Takiwaki2012, Hanke2013, Melson2015}. The results of these simulations indicate that some multi-dimensional hydrodynamic instabilities, such as convective motion \citep[e.g.,][]{Herant1994} and the standing-accretion-shock-instability \citep[SASI; e.g.,][]{Foglizzo2002, Blondin2003}, might play a crucial role in producing energetic explosions, because the random movement of the ejecta gas due to the instabilities can increase the neutrino heating efficiency in the gain region. Additional effects to enhance some instabilities have also been proposed to obtain realistic SN explosions. For instance, effects from general relativity, rotation, magnetic fields, and inhomogeneities in the progenitor core. The asymmetries resulting from the instabilities in the inner core might introduce Rayleigh-Taylor (RT) instability at the C+O/He and He/H composition interfaces and create global asymmetries in the ejecta \citep[e.g.,][]{Wongwathanarat2015}.

Investigating explosion geometries of SNe can provide insights on the explosion mechanism. In particular, it is critically important to clarify relations between the explosion asphericity and the SN properties (e.g., explosion energy, ejecta mass and progenitor radius). The proposed explosion scenarios can be tested by comparing these relations with those from theoretical simulations. 
%
%In fact, there are some observational implications for aspherical geometries in some SN explosions, from analysis of the shapes of spectral lines at the nebular phase (e.g., Elmhamdi et al. 2003; Chugai 2007; Maguire et al. 2012; Utrobin \& Chugai 2017; Andrews et al. 2019).
%Elmhamdi et al. (2003) pointed out the aspherical distribution of $^{56}$Ni in SN 1999em from the analysis of H$alpha$ and He I lines at the nebular phase.
%Chugai (2007) demonstrated that the double-peak shape of H$\alpha$ of SN 2004dj at the nebular phase can be explained by asymmetric disctribution of $^{56}$Ni.
%Maguire et al. (2012) line profiles => mixing
%Utrobin \& Chugai (2017) H$\alpha$ and LC at the radioactive tail
%Andrews et al. (2019)  multi-peaked emission lines of H$\alpha$ and [O I]
%However, since such late-time spectroscopic observations are difficult, only several targets have been studied in the point of the explosion geometry. In addition, there are many uncertainties in the processes of deriving the explosion structure quantitatively, i.e., in the processes of converting the shape of H$\alpha$ to the distribution of $^{56}$Ni.
%
The most reliable and direct way for investigating the geometry of a spatially unresolved SN is polarimetry. For an SN with a spherically symmetric photosphere, no polarization is expected due to cancellation of the orthogonal Stokes vectors. Thus, the detection of a non-null continuum polarization provides direct and independent evidence for an overall asphericity.

Hydrogen-rich CCSNe (hereafter Type~II SNe) are classified into two subtypes, Type~IIP and IIL, based on the light-curve shapes. Type~IIP SNe are the most common class of CCSNe \citep[$\sim 50$ \% of all CC events; e.g.,][]{Li2011}. They show constant optical brightness until $\sim 100$ days after explosion (the photospheric phase), which is followed by an exponential flux decline (the tail phase) after a sudden drop \citep[e.g.,][]{Anderson2014, Faran2014a, Sanders2015, Valenti2016}. In general, a low level of polarization ($\sim$0.1 \%) has been measured during the photospheric phase, while a rapid increase of continuum polarization ($\sim 1.0$ \%) is seen at the beginning of the tail phase in some Type~IIP SNe \citep[e.g.,][]{Leonard2001, Leonard2006, Chornock2010, Kumar2016}. This polarimetric feature can be explained by an asymmetric helium core being revealed when the outer spherical hydrogen envelope becomes optically thin as indicated by the light curve falling off the plateau \citep[e.g.,][]{Leonard2006}. On the other hand, \citet[][]{Dessart2011} demonstrated that the small continuum polarization during the photospheric phase can be, in some cases, reproduced by optical-depth effects irrespective of the magnitude of the asphericity of the hydrogen envelope, based on the polarimetric calculations from axially symmetric SN ejecta artificially constructed using 1-dimensional inputs from non-local-thermodynamic-equilibrium radiative-transfer simulations. Although only a few Type~IIP SNe have been observed in polarimetry with a sufficient temporal coverage, this high polarization is regarded as supporting evidence for the asymmetric explosions suggested by some numerical simulations \citep[e.g.,][]{Janka2012}. Recently, \citet[][]{Nagao2019} found an unprecedented, highly-extended aspherical explosion in SN~2017gmr indicated by an early rise of polarization, hence clearly showing that asymmetries are present not only in the helium core but also in a substantial portion of the hydrogen envelope. This demonstrates the existence of an intrinsic diversity in the aspherical structure of Type~IIP SN explosions.

Type~IIL SNe are a relatively rare subtype of Type~II SNe, which show a linear decline in their brightness during the photospheric phase \citep[e.g.,][]{Faran2014b}. There has not been many Type~IIL SNe observed with polarimetry. \citet[][]{Nagao2021} reported that a low polarization has been observed during the entire evolution from the photospheric to the tail phases in SN~2017ahn, while SN~2013ej showed a high polarization from the beginning of the photospheric phase. Based on the polarization analysis of SNe~2017ahn and 2013ej, \citet[][]{Nagao2021} claim that there are at least two subtypes of Type~IIL SNe in terms of their explosion schemes: some SNe may come from different explosions than Type~IIP SNe (a spherical explosion; e.g., SN~2017ahn) and others from similar explosions (an aspherical explosion; e.g., SN~2013ej).

In this study, we investigate the polarimetric properties of 15 Type~II SNe using unpublished and published spectropolarimetric data obtained with the Very Large Telescope (VLT) and other telescopes. We have divided the analysis into two papers. In Nagao et al. (2023; hereafter Paper~I), we present the full description of our SN sample, the observations, data reduction techniques, and our method to estimate the interstellar polarization (ISP). We also investigate the properties of the ISP components of 11 well-observed SNe from our sample in Paper~I.
Here, we discuss the properties of the intrinsic SN polarization of 15 SNe from our SN sample, their photometric and spectroscopic properties, and the relations between these properties, with the aim of exploring possible relations between the explosion asphericity and the explosion physics. Throughout this work, the classification between Type IIP and IIL is based on their LC shapes, i.e., quantified by the value of the decline rate during the photospheric phase ($s_{p}$; see Section~\ref{subsec:photo}). In the following sections, we show polarimetric (Section~\ref{sec:pol}), photometric and spectroscopic (Section~\ref{sec:photo_spec}) properties of our sample. In Section~\ref{sec:correlation}, we discuss relations between the polarimetric properties and the other observational properties of the SNe. Our conclusions are finally presented in Section~\ref{sec:conclusions}.

\section{The properties of the intrinsic SN polarization} \label{sec:pol}

\begin{figure*}
    \includegraphics[width=\columnwidth]{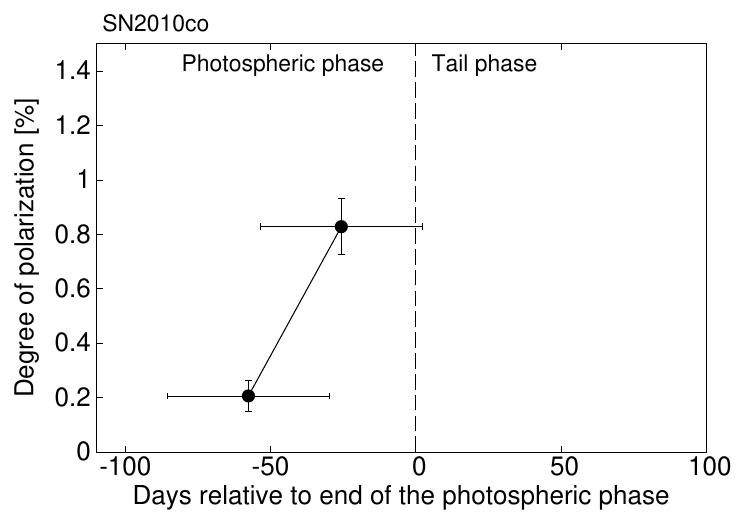}
    \includegraphics[width=\columnwidth]{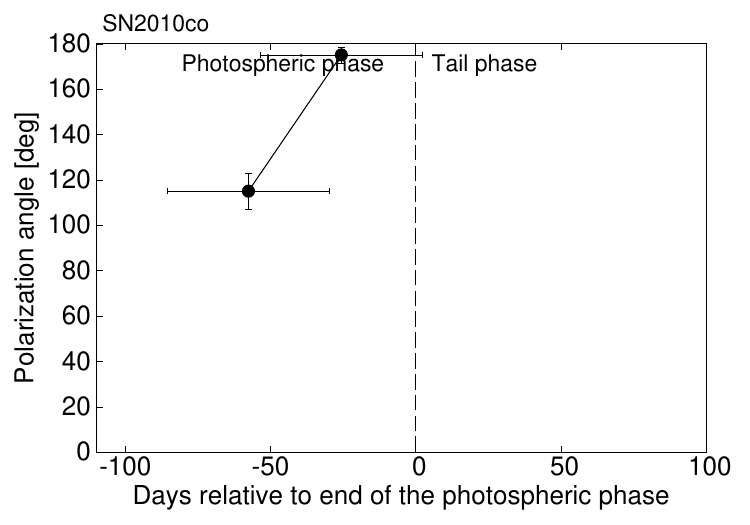}
    \caption{Polarization degree and angle of the continuum polarization in SN~2010co.}
    \label{fig:pol_10co}
\end{figure*}

\begin{figure*}
    \includegraphics[width=\columnwidth]{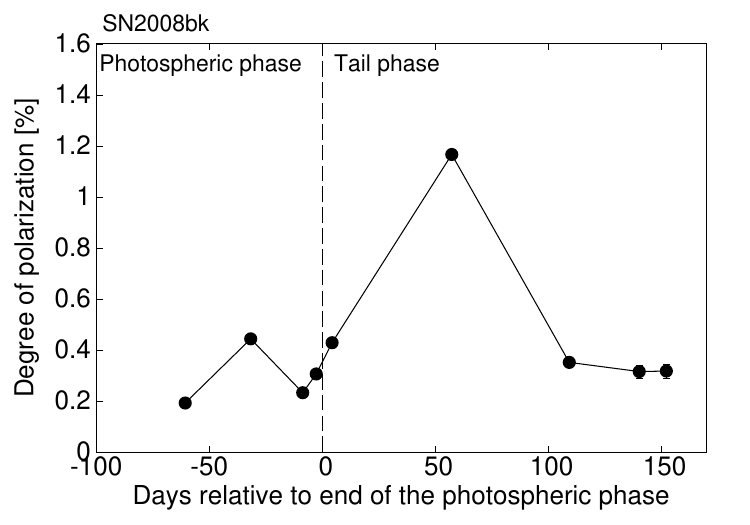}
    \includegraphics[width=\columnwidth]{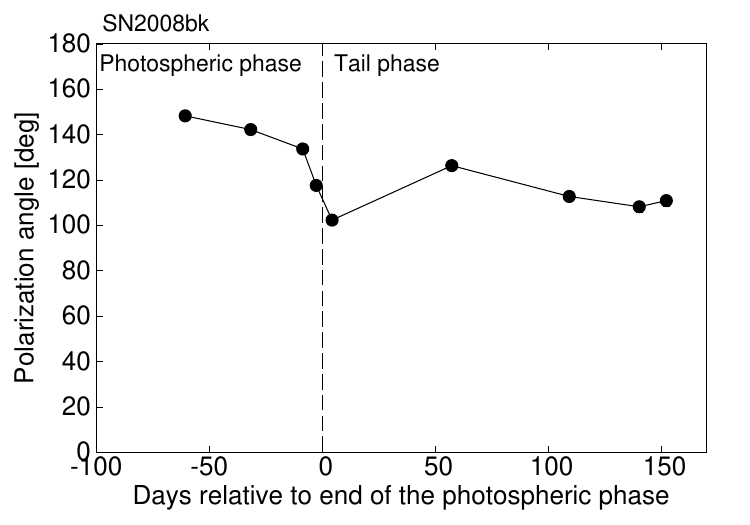}
    \caption{Polarization degree and angle of the continuum polarization in SN~2008bk.}
    \label{fig:pol_08bk}
\end{figure*}

\begin{figure*}
    \includegraphics[width=\columnwidth]{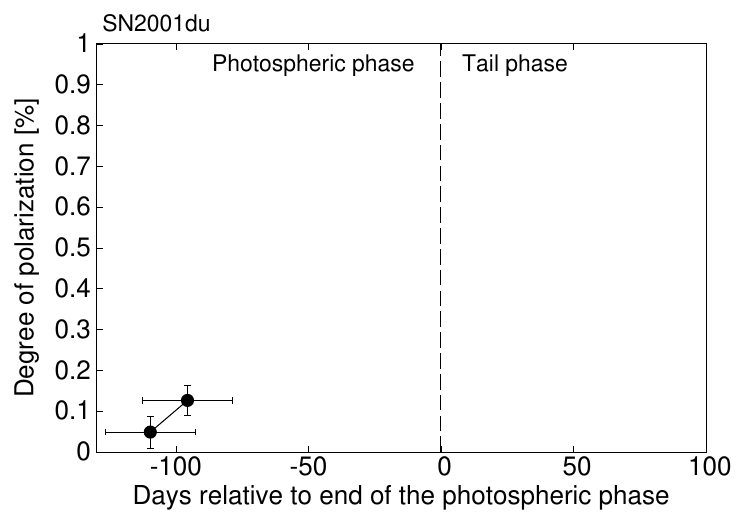}
    \includegraphics[width=\columnwidth]{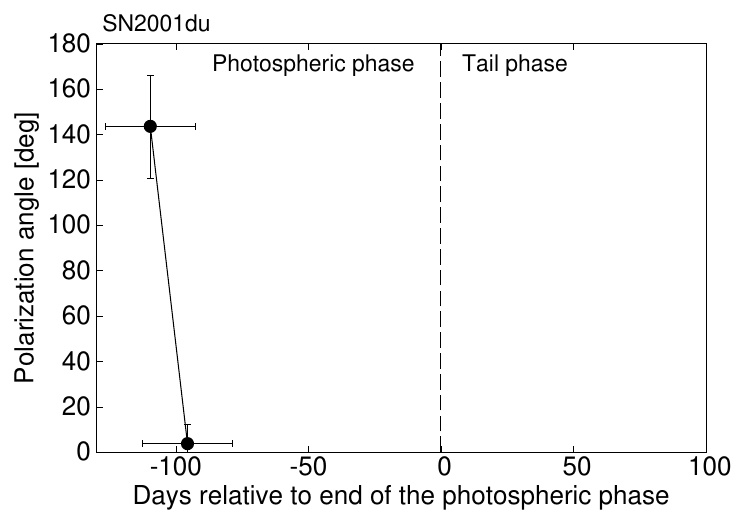}
    \caption{Polarization degree and angle of the continuum polarization in SN~2001du.}
    \label{fig:pol_01du}
\end{figure*}

\begin{figure*}
    \includegraphics[width=\columnwidth]{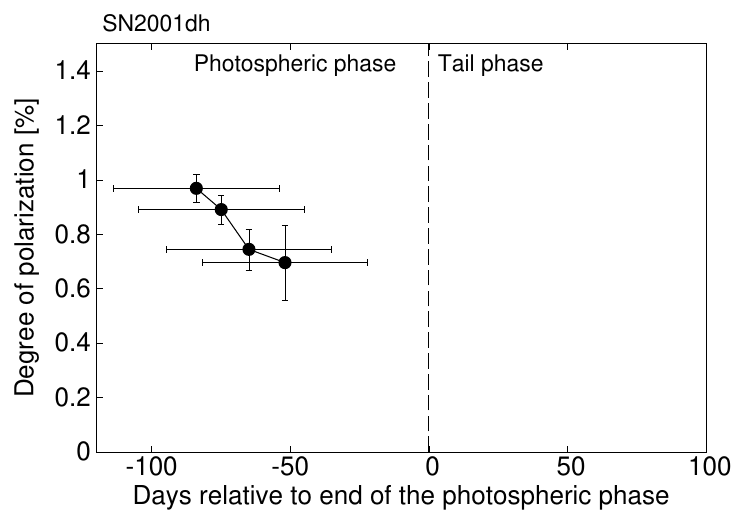}
    \includegraphics[width=\columnwidth]{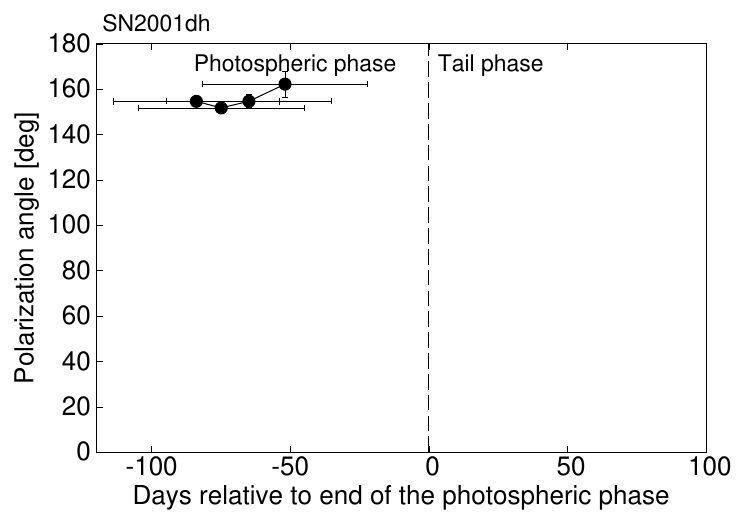}
    \caption{Polarization degree and angle of the continuum polarization in SN~2001dh.}
    \label{fig:pol_01dh}
\end{figure*}

In this section, we discuss the continuum polarization in the intrinsic SN polarization, which is derived by subtracting the ISP component from the observed polarization spectrum. The full description of our SN sample, the observations, the data reduction techniques, and the method to estimate the interstellar polarization are presented in Paper~I. 
As for the VLT sample, we calculate the continuum polarization in the same way with \citet[][]{Chornock2010}, \citet[][]{Nagao2019} and \citet[][]{Nagao2021}. We independently take the sums of the ordinary fluxes and the extraordinary fluxes within the wavelength ranges 6800-7200 {\AA} as well as 7820-8140 {\AA}, using the reduced 50{\AA}-binned polarization spectra (see Section~3.1 in Paper~I). From these averaged fluxes of the ordinary and extra-ordinary beams, we calculate the Stokes $Q$ and $U$ values and thereby the polarization degree and angle for the continuum polarization. The derived values are discussed in Section~\ref{subsec:individual_pol} and shown in Figures~\ref{fig:pol_17gmr}-\ref{fig:pol_01dh}.
It is noted that, for SNe~1999em, 2004dj, 2006ov and 2007aa, we have instead adopted the derived values for the ISP and the intrinsic SN polarization from the literature \citep[][]{Leonard2001, Leonard2006, Chornock2010}. As the continuum polarization of SN~1999em, we adopted the $V$-band polarization values that were measured by calculating the debiased, flux-weighted averages of $q$ and $u$ over the intervals 6050-6950 {\AA} for the first epoch and 5050-5950 {\AA} for the other epochs \citep[][]{Leonard2001}. For SN~2004dj, we adopted the continuum polarization values that were derived as the median values of the Stokes parameters over the interval 6800-8200 {\AA} \citep[][]{Leonard2006}. The continuum polarization of SNe 2006ov and 2007aa were derived in the same way with the one in this paper.

In addition to the ISP, there are two possible origins for the continuum polarization observed in Type~II SNe: electron scattering in an aspherical photosphere (the aspherical-photosphere origin) and dust scattering in aspherically distributed circumstellar material \citep[the dust scattering origin; see, e.g.,][and references therein]{Nagao2019, Nagao2021}. In the case of the aspherical-photosphere origin the polarization is wavelength independent, while in the dust scattering origin some wavelength dependence is expected \citep[see][for details]{Nagao2018}. Since all polarization spectra in our sample do not show any clear wavelength dependence (see Figures~\ref{fig:app_17gmr}-\ref{fig:app_01dh} in Appendix~\ref{sec:app_B}) at any phase, we conclude that the continuum polarization of all the SNe at all the epochs in our sample originate from aspherical structures of their photosphere. Such asphericities can be created by an aspherical explosion \citep[e.g., SN~2017gmr;][]{Nagao2019}, an ejecta interaction with aspherical CSM \citep[e.g., SN~2013ej;][]{Nagao2021} or other unknown processes.

\subsection{Individual SNe in the VLT sample} \label{subsec:individual_pol}

The VLT sample consists of the following 11 SNe: SNe~2017gmr, 2017ahn, 2013ej, 2012ec, 2012dh, 2012aw, 2010hv, 2010co, 2008bk, 2001du, and 2001dh. The individual objects are discussed in the following sub-sections.

\subsubsection{SN~2017gmr} \label{subsubsec:17gmr}
\begin{figure*}
    \includegraphics[width=\columnwidth]{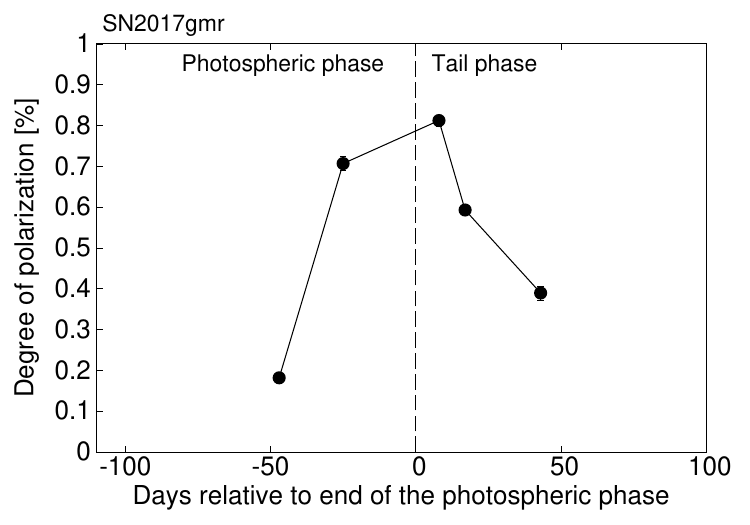}
    \includegraphics[width=\columnwidth]{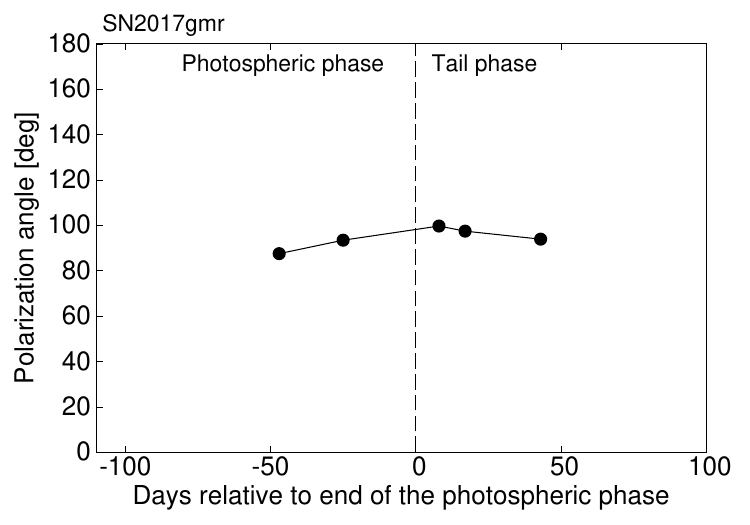}
    \caption{Polarization degree and angle of the continuum polarization in SN~2017gmr.}
    \label{fig:pol_17gmr}
\end{figure*}

The data of SN~2017gmr were already discussed in \citet[][]{Nagao2019}. As shown in that work, the intrinsic continuum polarization of SN~2017gmr shows high degrees ($\sim 0.8$ \%) with a constant angle ($\sim 95\degr$) from the middle of the photospheric phase to the beginning of the tail phase (see Figure~\ref{fig:pol_17gmr}). This indicates that the asymmetry is not confined to the helium core but extends to a significant part of the hydrogen envelope, and that the SN has a global aspherical structure from the helium core to the hydrogen envelope, e.g., a jet-like structure. Since the SN did not show any features due to a strong CSM interaction \citep[see][]{Andrews2019}, we interpret the origin of this polarization as an aspherical structure of the ejecta originated from an aspherical explosion \citep[see also discussions in][]{Nagao2019}. In addition, since the polarization evolution in the tail phase is similar to other Type~IIP SNe that are believed to be aspherical explosions \citep[$t^{-2}$ evolution, which can be interpreted as being due to geometrical dilution that reduces the optical depth of the ejecta; e.g., SN~2004dj][]{Leonard2006}, it is natural to interpret the origin as a common feature, i.e., aspherical explosions. This conclusion is supported by the asymmetric lines seen in the nebular spectra of SN~2017gmr \citep[e.g.,][]{Andrews2019, Utrobin2021}.

\subsubsection{SN~2017ahn} \label{subsubsec:17ahn}
\begin{figure*}
    \includegraphics[width=\columnwidth]{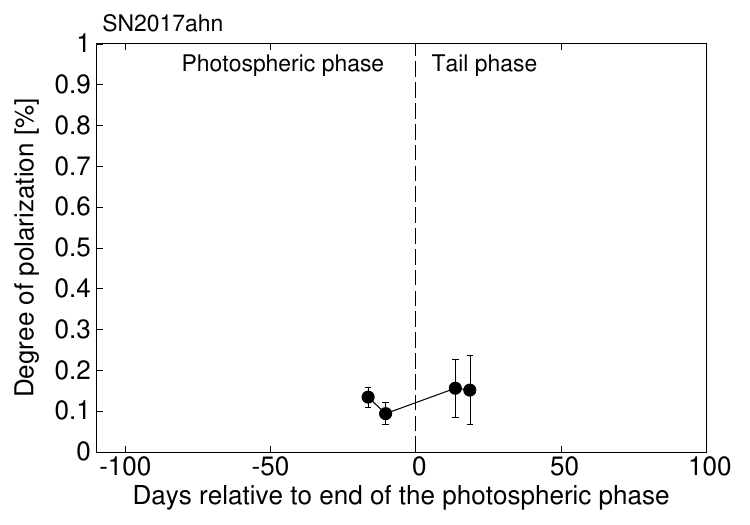}
    \includegraphics[width=\columnwidth]{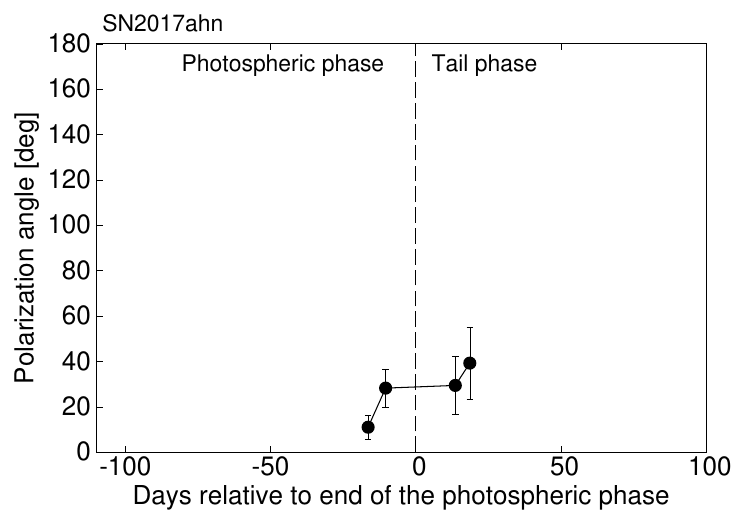}
    \caption{Polarization degree and angle of the continuum polarization in SN~2017ahn.
    }
    \label{fig:pol_17ahn}
\end{figure*}

The data of SN~2017ahn were already presented in \citet[][]{Nagao2021}. As discussed in that paper, the continuum polarization of this object remains low from the photospheric to tail phases (Figure~\ref{fig:pol_17ahn}), which implies a purely spherical explosion or a polar viewing angle of an aspherical explosion. We note that, at the first two epochs (and possibly also at the other epochs as well), the polarization might show a component in the bluer wavelengths ($\lambda < 6000$ \AA) with $\sim 0.5$ \% degree of polarization (see Figure~\ref{fig:app_17ahn}). This might originate from some line polarization, some dust scattering component or an incomplete ISP subtraction.

\subsubsection{SN~2013ej} \label{subsubsec:13ej}
\begin{figure*}
    \includegraphics[width=\columnwidth]{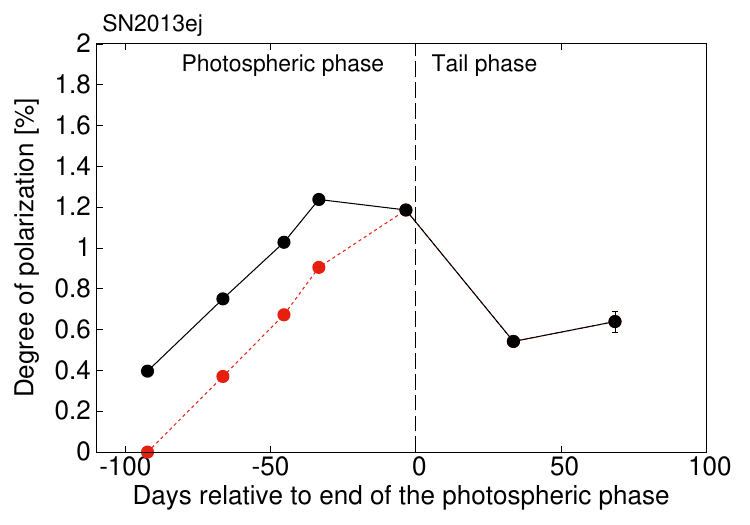}
    \includegraphics[width=\columnwidth]{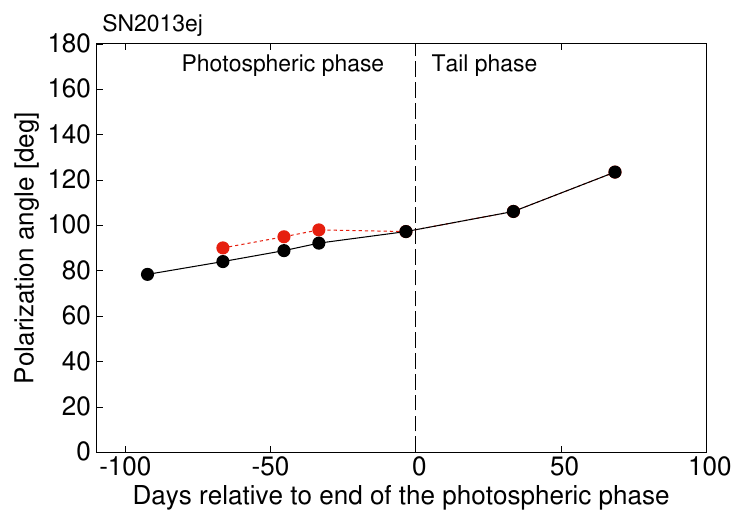}
    \caption{Polarization degree and angle of the continuum polarization in SN~2013ej. The explosion component is shown with a red color.}
    \label{fig:pol_13ej}
\end{figure*}

The data of SN~2013ej were already discussed in \citet[][]{Nagao2021}. The continuum polarization of this SN has two components with different angles, possibly consisting of an interaction component (originating from the aspherical CSM interaction) and an explosion component (arising from an aspherical explosion). Following the discussion in \citet[][]{Nagao2021}, we assume that the signal at the first epoch comes purely from the CSM interaction ($P=0.40$ \% and $\theta=78.4\degr$). The explosion component, which is derived by subtracting the interaction component from the observed polarization, is shown in Figure~\ref{fig:pol_13ej} and Table~\ref{tab:log_13ej}. Hereafter, we consider only the explosion component as the continuum polarization of SN~2013ej.

\subsubsection{SN~2012ec} \label{subsubsec:12ec}
\begin{figure*}
    \includegraphics[width=\columnwidth]{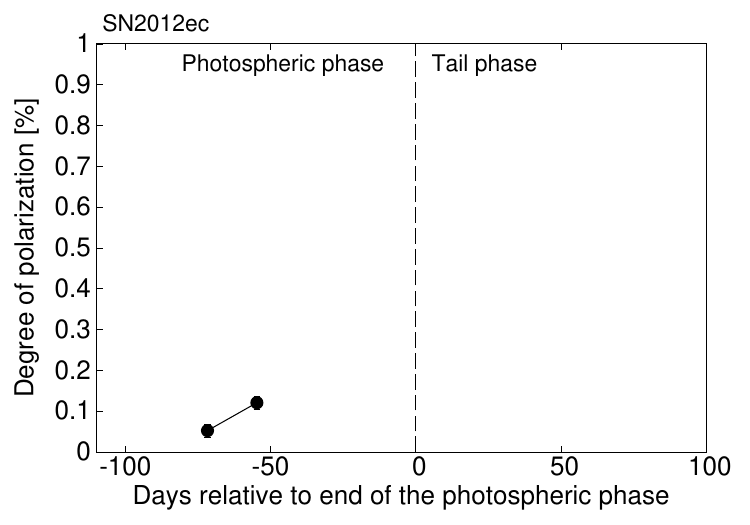}
    \includegraphics[width=\columnwidth]{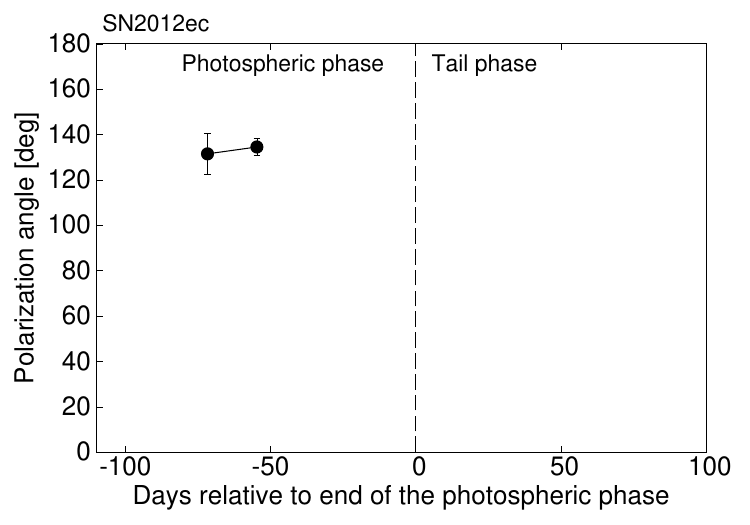}
    \caption{Polarization degree and angle of the continuum polarization in SN~2012ec.}
    \label{fig:pol_12ec}
\end{figure*}

This SN shows low polarization degrees at least until the latest observation (Phase -54.49 days; see Figure~\ref{fig:pol_12ec}). This implies that the shape of the photosphere is likely close to spherical, at least until this epoch.

\subsubsection{SN~2012dh} \label{subsubsec:12dh}
\begin{figure*}
    \includegraphics[width=\columnwidth]{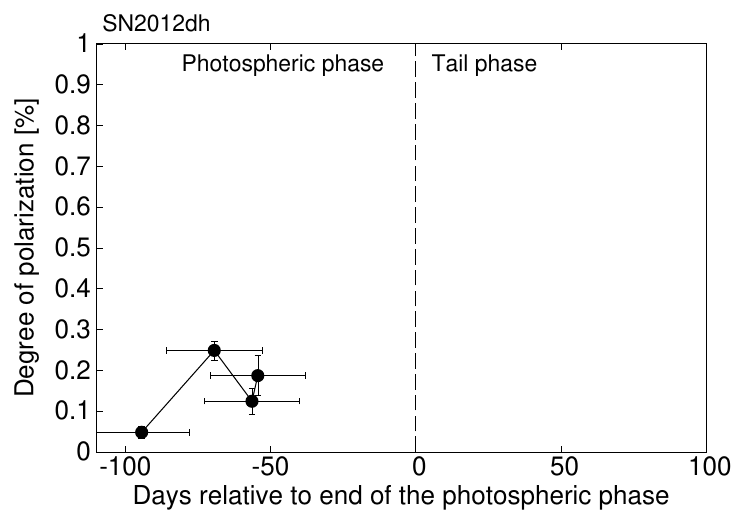}
    \includegraphics[width=\columnwidth]{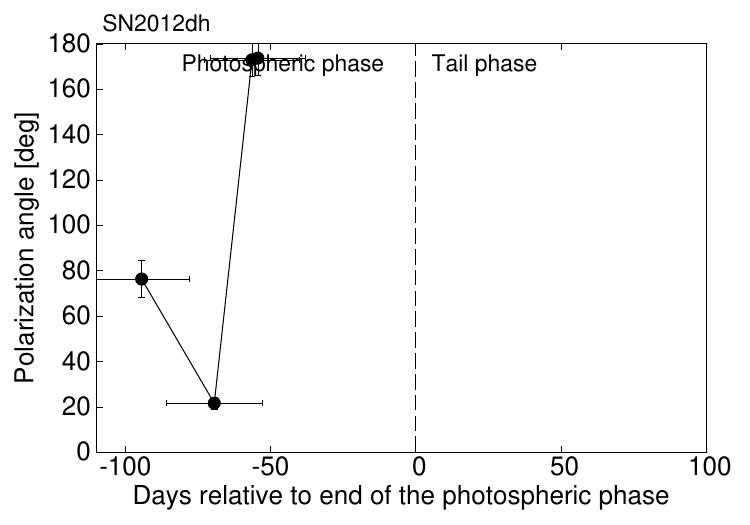}
    \caption{Polarization degree and angle of the continuum polarization in SN~2012dh.}
    \label{fig:pol_12dh}
\end{figure*}

SN~2012dh shows a relatively low polarization level ($P<0.3$ \%) until the epoch of the last observation (phase -54.37 $\pm$ 16.40 days), having erratic polarization angles (see Figure~\ref{fig:pol_12dh}). This implies that the shape of the photosphere is quite spherical, at least, until this phase. Here, the large errors bars in the phases originate from the large uncertainty of the end of the photospheric phase.

\subsubsection{SN~2012aw} \label{subsubsec:12aw}
\begin{figure*}
    \includegraphics[width=\columnwidth]{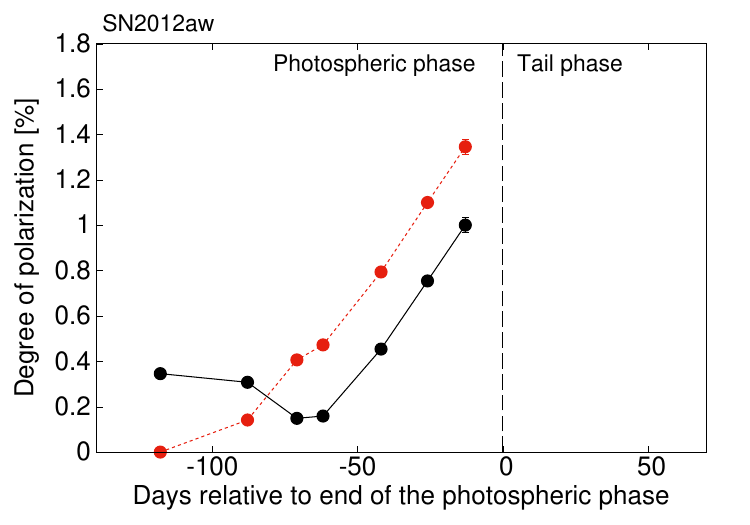}
    \includegraphics[width=\columnwidth]{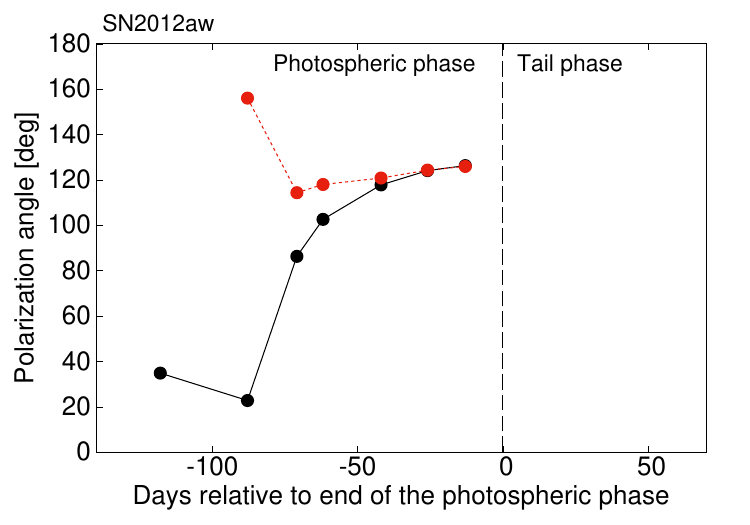}
    \caption{Polarization degree and angle of the continuum polarization in SN~2012aw. The aspherical explosion component is shown with a red color.}
    \label{fig:pol_12aw}
\end{figure*}

The data of SN~2012aw were already discussed in \citet[][]{Dessart2021}, even though they assumed slightly different values for the ISP (see Paper~I). As SN~2013ej, SN~2012aw has at least two components showing the polarization-angle change with time: a component with $\theta \sim 35\degr$ and a component with $\theta \sim 125\degr$ (see Figure~\ref{fig:pol_12aw}). The first two epochs appear to be dominated by the first component with a relatively constant polarization angle $P\sim 0.35$ \%, while the later epochs seem to be affected by the second time-evolving component with another constant polarization angle, $\theta \sim 125\degr$. This behavior of the continuum polarization is similar to that of SN~2013ej \citep[][]{Nagao2021}. As in that case, the first and second components might be due to a combination of an aspherical CSM interaction and an aspherical explosion, respectively. However, as opposed to SN~2013ej, SN~2012aw did not show strong interaction features in its photometric and spectroscopic properties at early phases. In addition, the two components are close to perpendicular. This might suggest the presence of two-component ejecta driven by a "jet" explosion (a collimated high-energy component and a canonical, bulk SN component).

Assuming the signals in the first epoch represents purely one component ($P\sim 0.35$ \% and $\theta=34.8\degr$), we derive the remaining component by subtracting this first component from the original polarization signals (see Figure~\ref{fig:pol_12aw}). In fact, the remaining component shows a relatively constant polarization angle ($\theta\sim 120\degr$) as well as a polarization rise similar to the behaviours of SNe~2017gmr and 2013ej. Thus, we assume that the remaining component is related to an aspherical explosion, as in other SNe, e.g., SNe~2017gmr and 2013ej.
We note that the angle of the second component at the second epoch differs from those at the other epochs. This may be due to the large uncertainty originating from the subtraction of similar vectors in the Q-U plane, which we do not take into account for the error estimation of the polarization angles.

The difference of the polarization angles of the first component and the explosion component is $\sim 90\degr$ for SN~2012aw, while $\sim 20\degr$ for SN~2013ej (see Tables~\ref{tab:log_13ej} and \ref{tab:log_12aw}). 
If the origin of the first component for SN~2012aw is the same as for SN~2013ej, which is supposed to come from an aspherical CSM interaction \citep[][]{Nagao2021}, this might indicate that the orientations of the aspherical structures of the explosion and the CSM distribution  are independent. In turn, this suggests that the two asphericities (the first and second polarization components) are produced by different mechanisms. Unlike SN~2013ej, SN~2012aw does not show any clear evidence for a strong CSM interaction in the LC and spectra during the photospheric phase. Therefore, the origin of the first component may be different in the two objects. The number of SNe showing such a component is still too small to allow any firm conclusion on its nature.
Hereafter, we consider only the second component (red points in Figure~\ref{fig:pol_12aw}) as the continuum polarization of SN~2012aw.

\subsubsection{SN~2010hv} \label{subsubsec:10hv}
\begin{figure*}
    \includegraphics[width=\columnwidth]{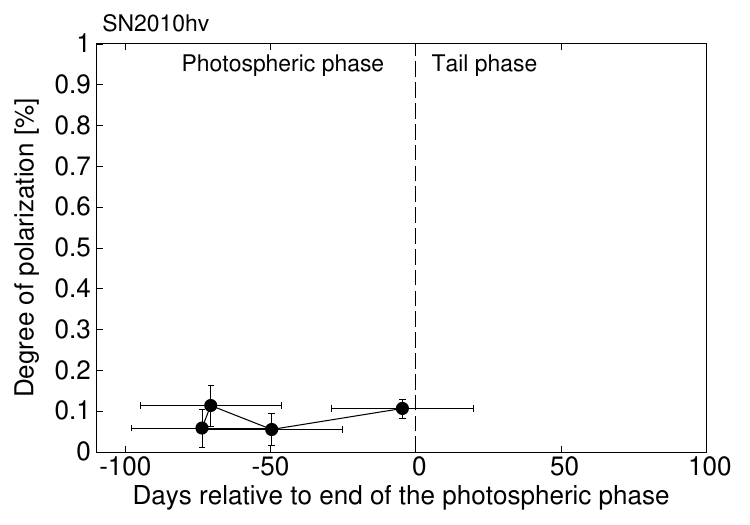}
    \includegraphics[width=\columnwidth]{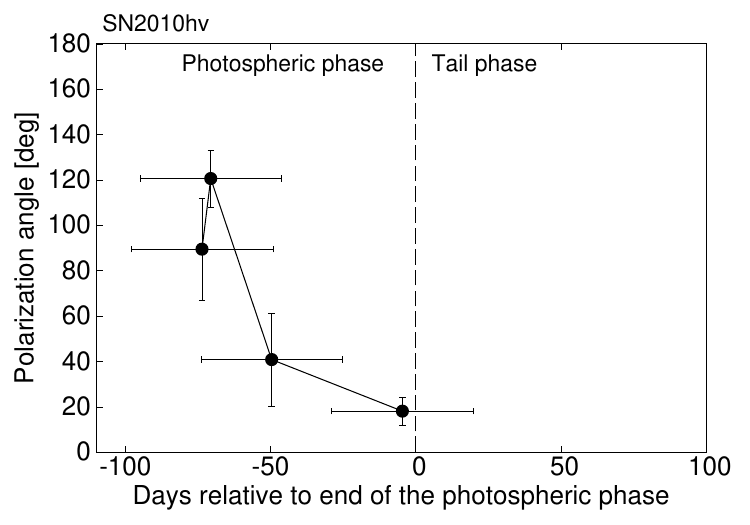}
    \caption{Polarization degree and angle of the continuum polarization in SN~2010hv.}
    \label{fig:pol_10hv}
\end{figure*}

SN~2010hv shows low polarization degrees during most parts of the photospheric phase (see Figure~\ref{fig:pol_10hv}), even though there is a large uncertainty in the phases. This implies that either most of the hydrogen envelope is quite spherical or that the viewing angle is close to polar for an aspherical structure.

\subsubsection{SN~2010co} \label{subsubsec:10co}
SN~2010co shows a polarization rise during the photospheric phase as SNe~2017gmr, 2013ej and 2012aw. This implies that this SN also has an aspherical structure extending to the hydrogen envelope.

\subsubsection{SN~2008bk} \label{subsubsec:08bk}
SN~2008bk shows relatively low polarization degrees during the photospheric phase, while it develops high polarization in the tail phase. It is not clear whether the timing of the polarization rise corresponds to the luminosity drop as in the other SNe, because that part of the polarimetric evolution (occurring between our fith and sixth epoch) is not covered by the available data. However, the general behavior is similar to that expected for the classical scenario \citep[e.g., SN~2004dj;][]{Leonard2006}: an aspherical He core surrounded by a spherical hydrogen envelope. Since the polarization degree around 60 days after the luminosity drop is high ($P\sim 1.17$ \%) compared with the other SNe in our sample (see Figure~\ref{fig:cont_pol}), the actual polarization peak might be also high ($\sim 2$ \%) at the timing just after the luminosity drop, if this SN also follows the $t^{-2}$ evolution. This might imply a very aspherical He core or just be due to the viewing angle effect of a similar spherical core in our sample.

\subsubsection{SN~2001du} \label{subsubsec:01du}
Although the data of SN~2001du are rather scanty, this object shows low polarization degrees at least at the start of the photospheric phase unlike SNe~2013ej and 2012aw. This behavior is seen in most Type~II SNe.

\subsubsection{SN~2001dh} \label{subsubsec:01dh}
The continuum polarization of SN~2001dh shows a very different behavior with respect to those of all the other SNe that have been observed so far, in that it displays a component with a single polarization angle ($\theta \sim 160\degr$), which declines from the $\sim 1$\% level during the photospheric phase. The 1\% of the continuum polarization at the start of the photospheric phase is the highest level among all observed Type~II SNe ever. In addition, the marked decline has not been observed in any other SNe. Since it is difficult for an aspherical explosion to create some aspherical structure only in the outermost layers, this polarization decline might be due to cancellation of polarization components by multiple aspherical structures as in SNe~2013ej and 2012aw.

\subsection{Diversity in the continuum polarization}

\begin{figure*}
    \includegraphics[width=2\columnwidth]{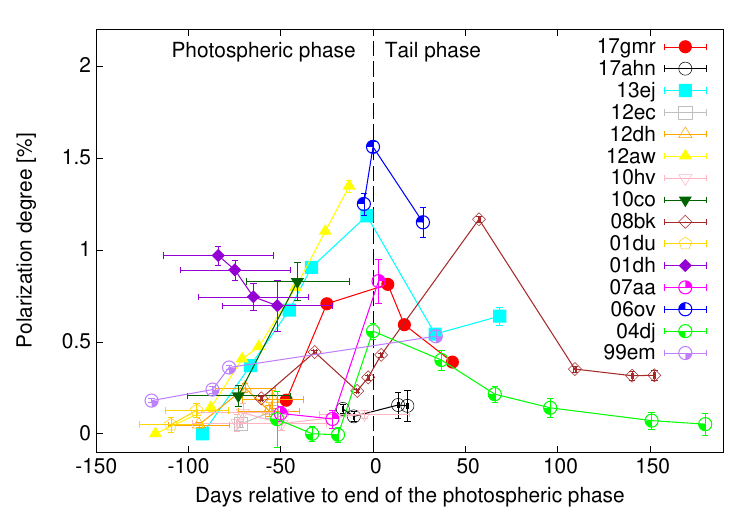}
    \caption{Time evolution of the continuum polarization of our SN sample. Here, for SNe~2013ej and 2012aw, the explosion components after the subtraction of the interaction components are plotted.}
    \label{fig:cont_pol}
\end{figure*}

Figure~\ref{fig:cont_pol} shows the time evolution of the continuum polarization of the SNe in our sample. All the SNe, with the only exception of SN~2017ahn, show relatively high polarization degrees at some epoch ($\sim 1$ \%), although there is large diversity in the temporal evolution. This supports that an aspherical structure is a common feature in Type~II SNe \citep[e.g.,][]{WangWheeler2008}. 

A polarization level as large as $\sim 1$ \% corresponds to an asphericity with a minimum axis ratio of 1.2:1.0 in the electron-scattering ellipsoidal atmosphere model by \citet[][]{Hoflich1991}, even though the peak polarization degree depends on the viewing angle. Given its behavior, SN~2017ahn might be an exceptionally spherical explosion, or a common aspherical structure viewed along the polar axis. In general, the timing of the polarization rise does not depend on the viewing angle unless we observe it somehow away from an axis of an aspherical structure. Thus, the different timings of the polarization rise mean that the extent of an aspherical structure in the hydrogen envelope is different for different SNe. 
As discussed above, SN~2001dh shows a behavior that is very different with respect to the others. This might imply a multi-directional explosion or an external factor, like an aspherical CSM interaction.

The observed diversity in the continuum polarization points to diverse explosion geometries in Type~II SNe, which can be categorized as follows:

\begin{itemize}
\item {\bf Group~1} (SNe~2008bk, 2007aa, 2006ov and 2004dj): this group includes SNe which show low polarization degrees ($<0.5$ \%) at the photospheric phase and a polarization rise ($\sim 1$ \%) at the photospheric to tail transitional phase. This is a classical picture of the continuum polarization in Type~II SNe \citep[e.g.,][]{Leonard2006}. This is explained in terms of an aspherical helium core enshrouded within a spherical hydrogen envelope. We note that SN~2006ov does not have observations during the photospheric phase, so it is not clear whether the polarization is low during the photospheric phase. However, it shows a clear polarization rise at the transitional phase, and thus may belong to this group.

\item {\bf Group~2} (SNe~2017gmr, 2013ej, 2012aw and 2010co): objects in this group show an increase of the polarization ($>0.5$ \%) already during the photospheric phase. This early rise implies that SNe have a very extended aspherical structure: asymmetries are not confined to the helium core but extend to a significant part of the hydrogen envelope.

\item {\bf Group~3} (SN~2017ahn): SNe in this group show low polarization at all times, from photospheric to tail phases. Since the properties of SN~2017ahn deviate from those of the other Type~II SNe \citep[][]{Tartaglia2021}, this might be generated by an explosion mechanism which is different from the majority of the explosions. Alternatively, this SN might be a similar explosion but, by chance, it was viewed along the axis of the aspherical structure. Since there is only such a case, it is difficult to draw any firm conclusion. Future larger samples may allow one to disentangle between possible options.
\end{itemize}

The polarization behaviour of SN~2001dh does not conform to any of the above groups. Since the observational data are very limited, we cannot conclude whether the high polarization originates just from an aspherical CSM interaction like SN~2013ej (belonging to one of above groups), or its explosion geometry is really aspherical to the outer ejecta edge (therefore belonging to a new group). As for the remaining SNe~2012ec, 2012dh, 2010hv, 2001du, 2001dh and 1999em, the data are not sufficient to decide to which of the above groups they belong to.

In the tail phase, the continuum polarization seems to have a behavior that is similar for all objects, i.e., a decline that approximately follows the so-called $t^{-2}$ evolution \citep[see][]{Leonard2006, Nagao2019}, which corresponds to the optical depth decreasing in an optically-thin, homologously expanding material. Although the number of SNe that have observations in the tail phase is limited, this indicates that the inner helium cores of Type~II SNe are generally aspherical. This implication is consistent with the polarimetric observations of stripped-envelope SNe \citep[e.g.,][]{Patat2001, Kawabata2002, Tanaka2009, Maund2009, Reilly2016}.

SNe~2013ej and 2017gmr, which have data both in the photospheric and in the tail phases, show a constant polarization angle throughout the observations. This implies a global aspherical structure from the inner to outer parts of the ejecta, which could be produced, for instance, by a collimated explosion.

\subsection{Characteristic value for the continuum polarization}

\begin{figure*}
    \includegraphics[width=\columnwidth]{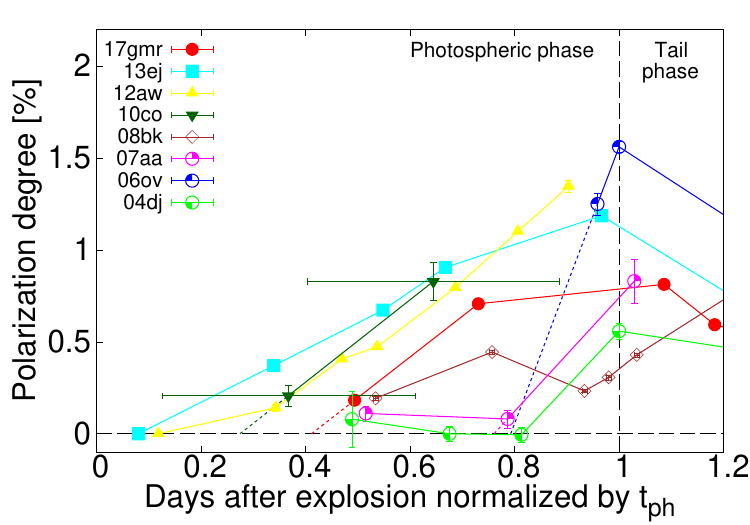}
    \includegraphics[width=\columnwidth]{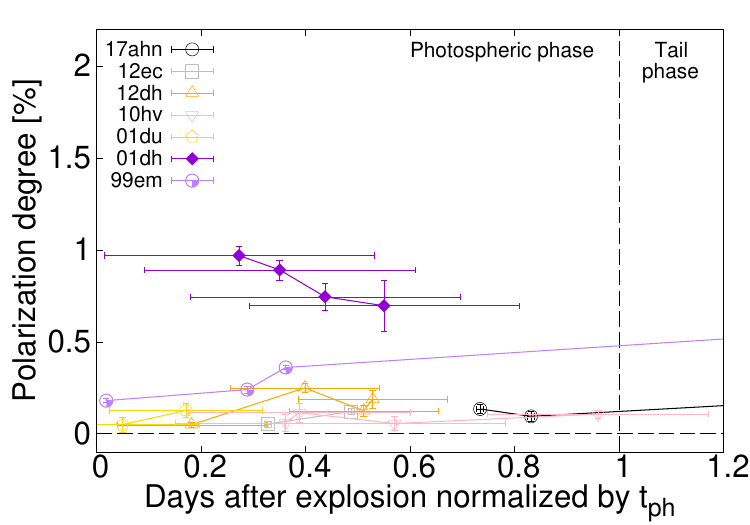}
    \caption{Same as Figure~\ref{fig:cont_pol} but with the time after explosion normalized by the length of the photospheric phase, for SNe in Groups 1 and 2 (left panel) and for the other SNe (right panel). The horizontal dashed lines express zero degree of polarization. The dotted lines are the extrapolation of the first two points that start to show an increase of the polarization degree, whose intersection with the dashed line ($P=0$) is defined as the timing of the polarization rise ($t_{\rm{pol}}$).}
    \label{fig:norm_cont_pol}
\end{figure*}

In general, the polarization of Type~II SNe shows low degrees at early photospheric phases, and then it increases to a $\sim 1.0\%$ level at a certain time during the photospheric phase: some SNe show an early rise, while in others this occurs just at the end of the photospheric phase (see Figure~\ref{fig:cont_pol}). The polarization degree depends on the shape and asymmetry of the photosphere and also on the viewing angle. On the other hand, the time of the polarization rise corresponds to the time when the photosphere recedes back to the outer edge of the aspherical structure, and it is therefore linked to the extent of this structure. Since the mass of the hydrogen envelope of Type~II SNe progenitors SNe varies from object to object, to allow for a meaningful comparison, we normalise the time of the polarization rise to the length of the photospheric phase, and we adopt the time of the polarization rise as a characteristic property of the continuum polarization ($t_{\rm{pol}}$). In practice, we estimate this time by extrapolating using the first two points that start to show an increase of the polarization degree for the SNe in Group~1 and 2 (see the left panel of Figure~\ref{fig:norm_cont_pol}). For the other SNe, whose data do not show a clear polarization rise, we use the last data point in the photospheric phase as an upper limit.
The derived values of $t_{\rm{pol}}$ are shown in Table~\ref{tab:SN}, together with other parameters (see below).

\renewcommand{\arraystretch}{1.2}
\begin{table*}
\caption{The type, polarization group and characteristic polarimetric, photometric and spectroscopic parameters for the 15 SNe in our sample.}
\label{tab:SN}
$
  \begin{tabular}{lccccccc} \hline
     Name & SN Type & Polarization & $t_{\rm{pol}}$ & $M_{V}(t_{\rm{ph}}/2)$, $L(t_{\rm{\rm{ph}}}/2)$ & $v(t_{\rm{ph}}/2)$ & $M_{\rm{Ni}}$\\
    && group && [mag], $10^{42}$[erg/s] & $10^{3}$[km/s] & [M$_{\odot}$]\\ \hline\hline
    SN~2017gmr & IIP & 2 & 0.41 & -18.1$\pm$0.2, 5.6$\pm$1.1 & 5.6$\pm$0.1 & 0.269$\pm$0.015 \\ \hline
    SN~2017ahn & IIL & 3 & (0.83-1.00) & -17.3$\pm$0.4, 2.7$\pm$1.2 & (4.7$\pm$0.1) & 0.038$\pm$0.002 \\ \hline
    SN~2013ej & IIL & 2 & 0.08 & -16.7$\pm$0.2, 1.6$\pm$0.3 & 4.5$\pm$0.1 & 0.019$\pm$0.001 \\ \hline
    SN~2012ec & IIP & ? & (0.49-1.00) & -16.5$\pm$0.1, 1.3$\pm$0.1 & 3.8$\pm$0.1 & 0.034$\pm$0.002 \\ \hline
    SN~2012dh & IIL & ? & (0.39-1.00) & -16.3$\pm$0.5, 1.1$\pm$0.7 & (4.6$\pm$0.1) & 0.024$\pm$0.002\\ \hline
    SN~2012aw & IIP & 2 & 0.12 & -16.7$\pm$0.1, 1.6$\pm$0.1 & 3.6$\pm$0.1 & 0.061$\pm$0.004 \\ \hline
    SN~2010hv & II & 1 or 3 & (0.75-1.00) & - & (4.2$\pm$0.1) & - \\ \hline
    SN~2010co & II & 2 & 0.28 $\pm$ 0.14 & - & (4.6$\pm$0.1) & - \\ \hline
    SN~2008bk & IIP & 1 & (0.98-1.00) & -14.7$\pm$0.1, 0.25$\pm$0.02 & (1.9$\pm$0.1) & 0.007$\pm$0.001 \\ \hline
    SN~2007aa & IIP & 1 & 0.76 & -16.7$\pm$0.2, 1.6$\pm$0.3 & 3.1$\pm$0.1 & 0.042$\pm$0.012 \\ \hline 
    SN~2006ov & IIP & 1 & 0.79 & -14.8$\pm$0.4, 0.27$\pm$0.12 & (1.7$\pm$0.1) & 0.002$\pm$0.001 \\ \hline 
    SN~2004dj & IIP & 1 & 0.82 & -16.0$\pm$0.2, 0.81$\pm$0.26 & 2.9$\pm$0.1 & 0.016$\pm$0.001 \\ \hline
    SN~2001du & II & ? & (0.05-1.00) & - & (5.8$\pm$0.1) & - \\ \hline
    SN~2001dh & II & 2? &  - & - & (3.2$\pm$0.1) & - \\ \hline
    SN~1999em & IIP & 1 or 2 & (0.36-1.00) & -16.7$\pm$0.2, 1.6$\pm$0.3 & 3.0$\pm$0.1 & 0.058$\pm$0.002 \\ \hline
  \end{tabular}
  $
\end{table*}

\section{Photometric and spectroscopic properties} \label{sec:photo_spec}

\subsection{Photometric properties} \label{subsec:photo}

\begin{figure*}
    \includegraphics[width=\columnwidth]{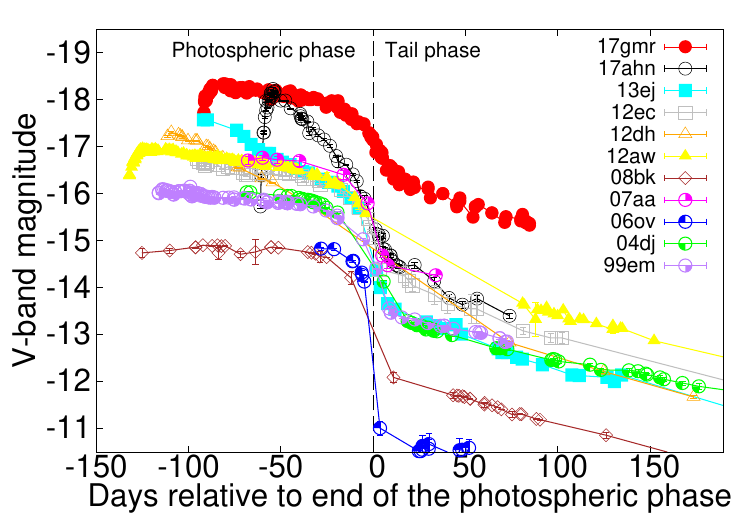}
    \includegraphics[width=\columnwidth]{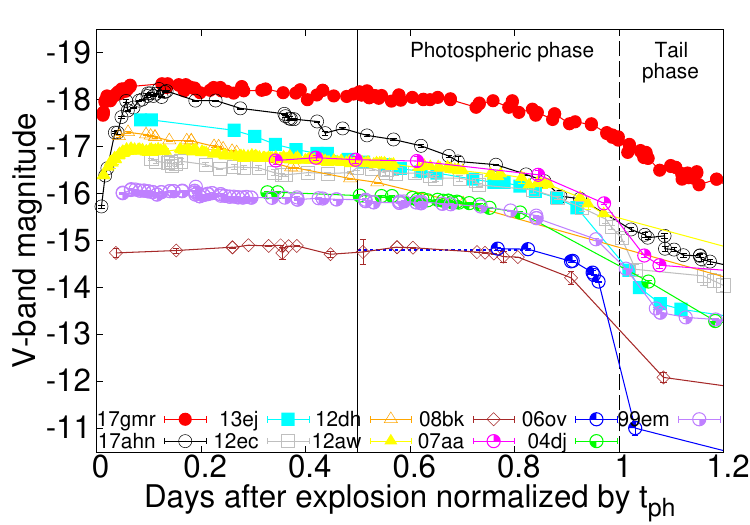}
    \caption{\textit{V}-band light curves of our sample (left panel) and the same light curves with the time after explosion nomalized by the length of the photospheric phase (right panel). The vertical line and dashed lines indicate the timing of the middle and end of the photospheric phase, respectively. The blue dotted line is the extrapolation of the first point.}
    \label{fig:LC}
\end{figure*}

\begin{figure}
    \includegraphics[width=\columnwidth]{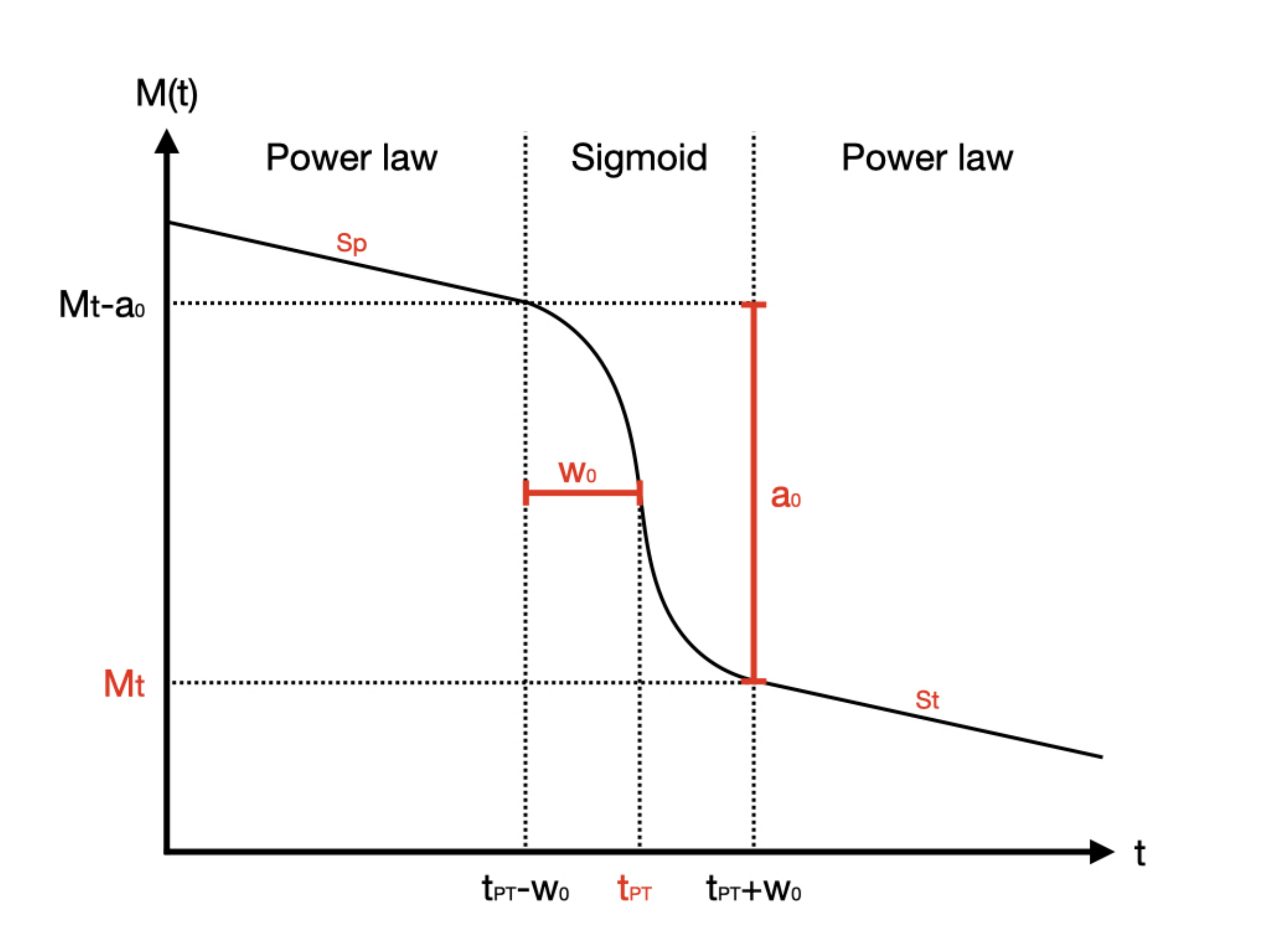}
    \caption{Schematics of the function used for the light curve fitting (see Eq.~\eqref{eq:FF})}
    \label{fig:FF}
\end{figure}

Figure~\ref{fig:LC} shows the light curves (LCs) of our SN sample, which are corrected for reddening (see Table~\ref{tab:SN} in Paper~I). The SNe have diverse properties in terms of brightness and duration of the photospheric phase. For characterizing the LC shapes, we fit the time evolution of the absolute magnitude ($M(t)$) with the following artificial function composed by two power law functions and a sigmoid function (see Figure~\ref{fig:FF}):

\begin{eqnarray}
&& M(t) = \frac{-a_{0}}{1+e^{\frac{t-t_{\rm{ph}}}{w_0}}} + M_{t} \nonumber\\
&& + \left\{ \begin{array}{ll}
    s_p \left[ t- \left( t_{\rm{ph}} - w_0 \right) \right] & (t \leq t_{\rm{ph}}-w_0) \\
    0 & (t_{\rm{ph}}-w_0 \leq t \leq t_{\rm{ph}}+w_0)\\
    s_t \left[ t- \left( t_{\rm{ph}} - w_0 \right) \right] & (t \geq t_{\rm{ph}}+w_0)
  \end{array} \right.
  \label{eq:FF}
\end{eqnarray}

where $M_{t}$, $a_{0}$, $t_{\rm{\rm{ph}}}$, $w_{0}$, $s_{\rm{p}}$ and $s_{\rm{t}}$ are free parameters. The best-fit values for these parameters are summarized in Table~\ref{tab:best_fit} for each SN.

As a characteristic value for the SN brightness we adopt the \textit{V}-band absolute magnitude in the middle of the photospheric phase ($M_{V}(t_{\rm{\rm{ph}}}/2)$). For deriving the typical value of the SN bolometric luminosity from the \textit{V}-band magnitude we simply assume a 6000 K black-body spectrum. The values of $L(t_{\rm{\rm{ph}}}/2)$ converted from $M(t_{\rm{\rm{ph}}}/2)$ are presented in Table~\ref{tab:SN}. 
For deriving the Ni mass for each SN, also summarized in Table~\ref{tab:SN}, we adopt the empirical method proposed by \citet[][]{Hamuy2003}, using the following equations:

\begin{equation}
    M_{\rm{Ni}} = (7.866 \times 10^{-44}) L_{t} \exp \left[ \frac{(t_{\rm{ph}}+w_{0})/(1+z) -6.1}{111.26} \right] \rm{M}_{\odot},
\end{equation}

where $L_{t}$ is the bolometric luminosity at $t=t_{\rm{ph}}+w_{0}$ which, in turn, is estimated using the following empirical equation by Hamuy (2001):

\begin{equation}
    \log_{10} L_{t} = \frac{-\left( M_{t} - A_{V} + BC \right) + 5 \log \left( 10 \rm{pc}\right) -8.14}{2.5}.
\end{equation}

Here, for the the bolometric correction ($BC$) we have adopted the value $BC=0.26 \pm 0.06$ during the tail phase (Hamuy 2001).

In this work, we have classified the SNe between Type IIP and IIL based on the values of $s_{p}$. The SNe with $s_{p} < 1$ are classified as Type~IIP, while the others are as Type~IIL. Here, the derived value of $s_{p}$ for SN~2006ov indicates it as a Type IIL SN in our criteria ($s_{p} > 1$; see Section~\ref{subsec:photo}). However, this is just due to the limited photometric data during the photospheric phase (see Section~\ref{subsec:photo}). The photometric and spectroscopic properties of SN~2006ov in \citet[][]{Chornock2010} suggest a class of Type~IIP. Thus, we classify it as a Type~IIP SN in this work. As for the SNe that do not have photometric observations (SNe~2010hv, 2010co, 2001du and 2001dh), we label them just as Type~II in this paper. The photometric classification of the SNe is shown in Table~\ref{tab:SN}.

\renewcommand{\arraystretch}{1.2}
\begin{table*}
\caption{Best-fit values from the LC fitting}
\label{tab:best_fit}
$
  \begin{tabular}{lcccccc} \hline
    Name & $t_{\rm{ph}}$ & $a_{0}$ & $w_{0}$ & $M_{t}$ & $s_{p}$ & $s_{t}$\\ \hline\hline
    SN~2017gmr & 92.92 $\pm$ 0.13 & 1.50 $\pm$ 0.01 & 5.58 $\pm$ 0.09 & -16.39 $\pm$ 0.01 & 0.56 $\pm$ 0.01 & 1.29 $\pm$ 0.02\\ \hline
    SN~2017ahn & 61.62 $\pm$ 0.41 & 1.65 $\pm$ 0.05 & 4.04 $\pm$ 0.22 & -14.59 $\pm$ 0.04 & 3.96 $\pm$ 0.06 & 1.75 $\pm$ 0.10\\ \hline
    SN~2013ej & 100.47 $\pm$ 0.08 & 2.38 $\pm$ 0.01 & 3.73 $\pm$ 0.06 & -13.47 $\pm$ 0.01 & 1.90 $\pm$ 0.02 & 1.17 $\pm$ 0.01\\ \hline
    SN~2012ec & 106.70 $\pm$ 0.67 & 2.22 $\pm$ 0.07 & 7.13 $\pm$ 0.56 & -14.01 $\pm$ 0.06 & 0.45 $\pm$ 0.04 & 1.11 $\pm$ 0.02\\ \hline
    SN~2012dh & (115.4 $\pm$ 16.4) & 1.53$^{+0.18}_{-0.17}$ & (5.8) & -13.53$^{+0.19}_{-0.22}$ & 2.18$^{+0.02}_{-0.06}$ & 1.16 $\pm$ 0.04\\ \hline
    SN~2012aw & 133.83 $\pm$ 0.33 & 2.01 $\pm$ 0.02 & 9.19 $\pm$ 0.32 & -14.35 $\pm$ 0.01 & 0.51 $\pm$ 0.01 & 1.01 $\pm$ 0.02\\ \hline
    SN~2008bk & 130.10 $\pm$ 0.65 & 2.82 $\pm$ 0.02 & 10.75 $\pm$ 0.26 & -12.05 $\pm$ 0.01 & 0.02 $\pm$ 0.01 & 1.06 $\pm$ 0.01\\ \hline
    SN~2007aa & 103.20 $\pm$ 1.22 & 2.10 $\pm$ 0.31 & 3.31 $\pm$ 0.76 & -14.34 $\pm$ 0.26 & 0.61 $\pm$ 0.24 & 0.15 $\pm$ 0.85\\ \hline 
    SN~2006ov & 121.37 $\pm$ 0.33 & 3.85 $\pm$ 0.07 & 2.08 $\pm$ 0.16 & -10.58 $\pm$ 0.06 & 1.78 $\pm$ 0.18 & 0.23 $\pm$ 0.09\\ \hline 
    SN~2004dj & 101.77 $\pm$ 0.52 & 2.54 $\pm$ 0.04 & 8.01 $\pm$ 0.31 & -13.22 $\pm$ 0.02 & 0.47 $\pm$ 0.06 & 0.80 $\pm$ 0.02\\ \hline 
    SN~1999em & 122.34 $\pm$ 0.12 & 2.17 $\pm$ 0.01 & 4.13 $\pm$ 0.09 & -14.15 $\pm$ 0.01 & 0.49 $\pm$ 0.01 & 0.65 $\pm$ 0.03\\ \hline
    \noalign{\smallskip}
    \multicolumn{7}{l}{Notes. The fitting function according to Equation~\ref{eq:FF} and Figure~\ref{fig:FF}.}
  \end{tabular}
  $
\end{table*}

\subsection{Spectroscopic properties}

\begin{figure*}
    \includegraphics[width=\columnwidth]{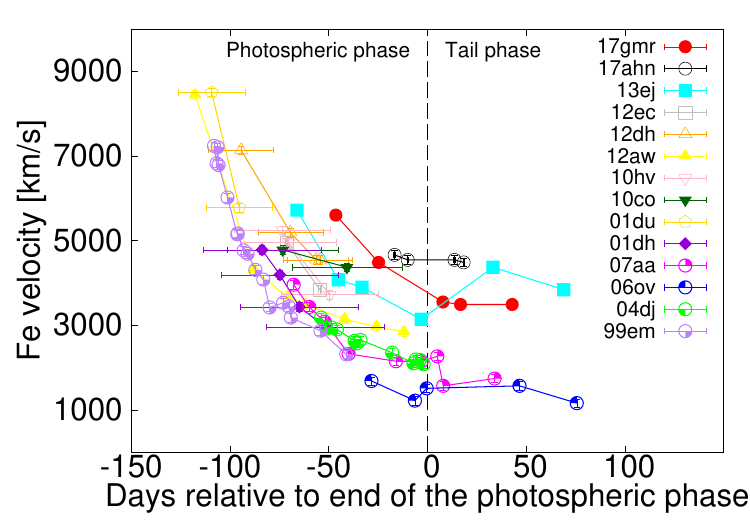}
    \includegraphics[width=\columnwidth]{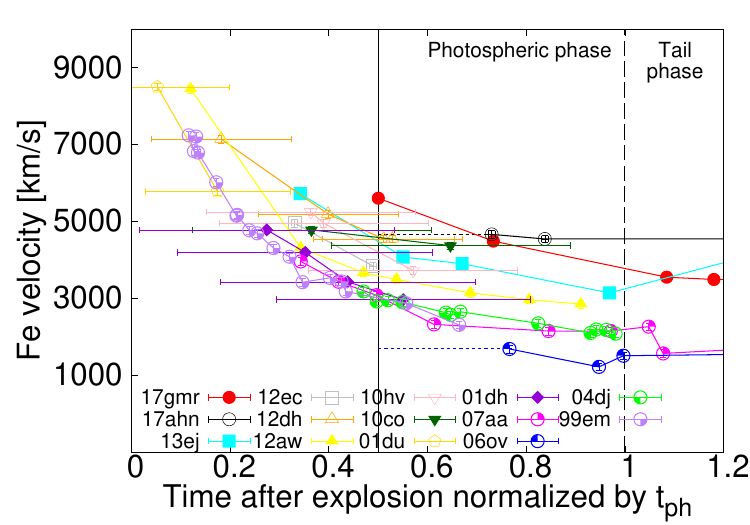}
    \caption{Fe~II velocities of our SN sample (left panel) and the velocities with the time after explosion nomalized by the length of the photospheric phase (right panel). The vertical lines are the same as in Figure~\ref{fig:LC}. The dotted lines are the extrapolation of the first point.}
    \label{fig:vel}
\end{figure*}

In this paper we do not discuss the spectroscopic properties of our sample. However, for the purposes of our analysis, we have extracted the expansion velocities. In particular, we have derived the ejecta velocity at the middle of the photospheric phase derived using the Fe II $\lambda 5169$ line ($v(t_{\rm{\rm{ph}}}/2)$), and used it as a characteristic value. The velocities were estimated from the absorption minima, using the spectra presented in Paper~I. For SNe~2004dj and 1999em, we adopted the Fe velocity derived by \citet[][]{Takats2012}. The uncertainties on the absorption minima position are evaluated from the spectral the dispersion (e.g., $\sim 3.3$ {\AA} pixel$^{-1}$ for G300V/FORS2/VLT). The derived velocities are shown in Figure~\ref{fig:vel} and the values of $v(t_{\rm{\rm{ph}}}/2)$ are provided in Table~\ref{tab:SN}.

\subsection{SN explosion properties} \label{sec:SN_explosion_properties}

We estimate the basic SN physical properties (explosion energy: $\widetilde{E}_{\rm{sn}}$, ejecta mass: $\widetilde{M}_{\rm{ej}}$, and pre-explosion progenitor radius: $\widetilde{R}_{\rm{p}}$) from the SN photometric and spectroscopic properties (typical luminosity: $\widetilde{L}_{\rm{sn}}$, typical duration: $\widetilde{t}_{\rm{sn}}$, typical ejecta velocity: $\widetilde{v}_{\rm{sn}}$, and the $^{56}$Ni mass: $M_{\rm{Ni}}$).
Here, we are not interested in deriving precise values for these physical quantities but rather in obtaining some relative scales for the quantities. Thus, we use a simple one-zone model of homologously expanding gas with radiative diffusion, as proposed by \citet[][]{Kasen2009}. This is characterised by the following equations:

\begin{eqnarray}
    \widetilde{M}_{\rm{ej}} &=& \frac{4 \pi c \sigma_{\rm{SB}} T_{I}^{4}}{\kappa_{\rm{es}}} \widetilde{t}_{\rm{sn}}^{4} \widetilde{L}_{\rm{sn}}^{-1} \widetilde{v}_{\rm{sn}}^{3}, \label{eq:eq4}\\
    \widetilde{E}_{\rm{sn}} &=& \frac{2 \pi c \sigma_{\rm{SB}} T_{I}^{4}}{\kappa_{\rm{es}}} \widetilde{t}_{\rm{sn}}^{4} \widetilde{L}_{\rm{sn}}^{-1} \widetilde{v}_{\rm{sn}}^{5}, \label{eq:eq5}\\
    \widetilde{R}_{\rm{p}} &=& \frac{\kappa_{\rm{es}}}{2 \pi c \sigma_{\rm{SB}} T_{I}^{4}} \widetilde{t}_{\rm{sn}}^{-2} \widetilde{L}_{\rm{sn}}^{2} \widetilde{v}_{\rm{sn}}^{-4} \left[ 1 - \widetilde{L}_{\rm{sn}}^{-1} \widetilde{t}_{\rm{sn}}^{-2} (E_{\rm{Ni}}t_{\rm{Ni}} +E_{\rm{Co}}t_{\rm{Co}}) \right], \label{eq:eq6}
\end{eqnarray}

where $E_{\rm{Ni}}\simeq0.6\times10^{50}(M_{\rm{Ni}}/M_{\odot})$ erg and $E_{\rm{Co}}\simeq1.2\times10^{50}(M_{\rm{Ni}}/M_{\odot})$ erg are the total energy by $^{56}$Ni and $^{56}$Co decay, and $t_{\rm{Ni}}\simeq8.8$ days and $t_{\rm{Co}}\simeq113$ days are the decay time. Here, we simply assume that $\widetilde{t}_{\rm{sn}}=t_{\rm{\rm{ph}}}$, $\widetilde{L}_{\rm{sn}}=L(t_{\rm{\rm{ph}}}/2)$ and $\widetilde{v}_{\rm{sn}}=v(t_{\rm{\rm{ph}}}/2)$.

We note that, since we assume a constant brightness by the radiative diffusion, we cannot apply this analysis to Type~IIL SNe in general. However, SN~2013ej is also proposed as a Type~IIP SN based on its photometric and spectroscopic properties \citep[e.g.,][]{Dhungana2016}. In fact, its LC is very similar to those of Type~IIP SNe except the early bump during the first $\sim 40$ days, which is understood as a result of a CSM interaction. Thus, we conducted this analysis for the Type~IIP SNe that have the photometric and spectroscopic information (SNe~2017gmr, 2012ec, 2012aw, 2008bk, 2007aa, 2006ov, 2004dj and 1999em) and the Type~IIL SN~2013ej.

The derived values for $\widetilde{E}_{\rm{sn}}$, $\widetilde{M}_{\rm{ej}}$ and $\widetilde{R}_{\rm{p}}$ are summarized in Table~\ref{tab:characteristic_values}.

\renewcommand{\arraystretch}{1.5}
\begin{table}
\caption{Characteristic SN explosion properties from the SN
photometric and spectroscopic properties.}
\label{tab:characteristic_values}
$
  \begin{tabular}{lccc} \hline
    Name & $\widetilde{M}_{\rm{ej}}$ & $\widetilde{E}_{\rm{sn}}$ & $\widetilde{R}_{\rm{p}}$\\
    & [M$_{\odot}$] & $10^{51}$ [erg/s] & [100$R_{\odot}$]\\ \hline\hline
    SN~2017gmr &  5.32$_{-1.13}^{+1.70}$ & 1.66$_{-0.40}^{+0.61}$ & 1.63$_{-1.63}^{+4.87}$ \\ \hline
    SN~2013ej & 13.20$_{-2.84}^{+4.21}$ & 2.66$_{-0.67}^{+1.00}$ & 2.37$_{-1.02}^{+1.43}$ \\ \hline
    SN~2012ec  & 12.45$_{-2.05}^{+2.50}$ & 1.79$_{-0.37}^{+0.47}$ & 2.11$_{-0.58}^{+0.75}$ \\ \hline
    SN~2012aw  & 21.28$_{-3.05}^{+3.61}$ & 2.74$_{-0.52}^{+0.65}$ & 2.63$_{-0.63}^{+0.81}$ \\ \hline
    SN~2008bk  & 17.88$_{-4.08}^{+5.25}$ & 0.64$_{-0.20}^{+0.28}$ & 0.98$_{-0.33}^{+0.49}$ \\ \hline
    SN~2007aa  & 4.81$_{-1.31}^{+2.01}$ & 0.46$_{-0.15}^{+0.23}$ & 7.38$_{-3.76}^{+5.89}$ \\ \hline 
    SN~2006ov  & 8.98$_{-3.85}^{+10.43}$ & 0.26$_{-0.13}^{+0.37}$ & 2.58$_{-1.99}^{+4.50}$ \\ \hline 
    SN~2004dj  & 7.35$_{-2.44}^{+4.88}$ & 0.61$_{-0.23}^{+0.48}$ & 2.93$_{-2.01}^{+3.63}$ \\ \hline
    SN~1999em  & 8.60$_{-2.08}^{+3.13}$ & 0.77$_{-0.22}^{+0.35}$ & 6.06$_{-3.08}^{+4.74}$ \\ \hline
    \noalign{\smallskip}
    \multicolumn{4}{l}{Notes. These values were estimated using Equations~\ref{eq:eq4}-\ref{eq:eq6}.}
  \end{tabular}
  $
\end{table}

\section{Relations between polarimetric and spectro-photometric properties} \label{sec:correlation}

In this section, we discuss relations between the polarimetric and photometric/spectroscopic parameters of eleven SNe that have both sets of observations (eight Type~IIP and three Type~IIL SNe).

\begin{figure*}
    \includegraphics[width=\columnwidth]{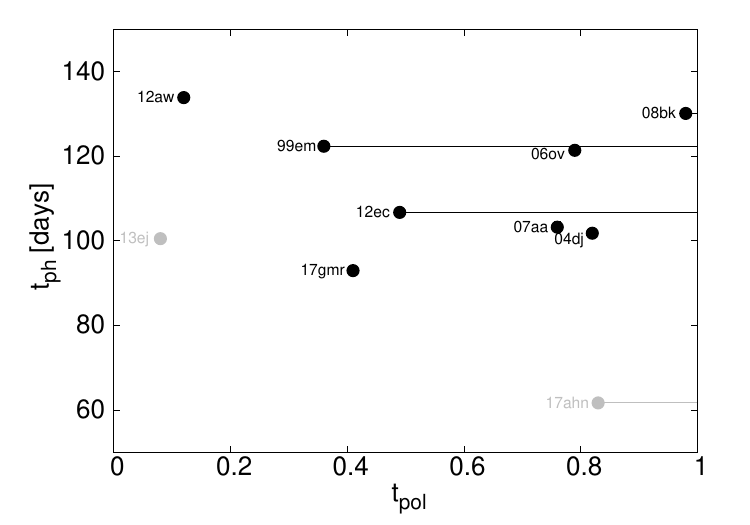}
    \includegraphics[width=\columnwidth]{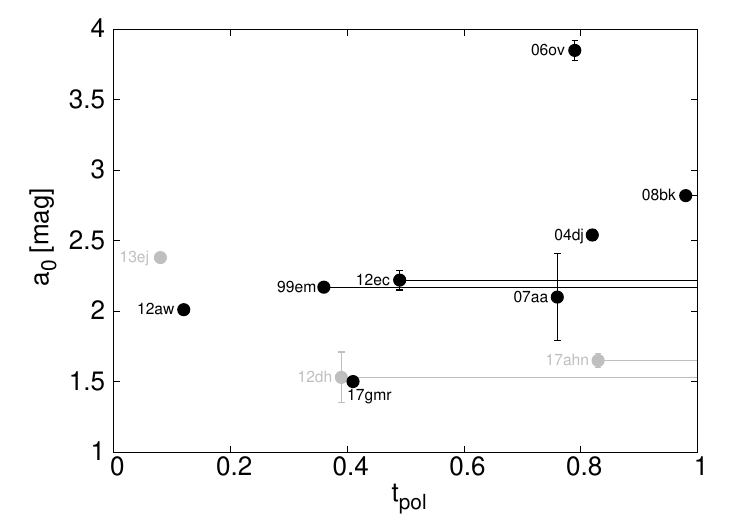}
    \includegraphics[width=\columnwidth]{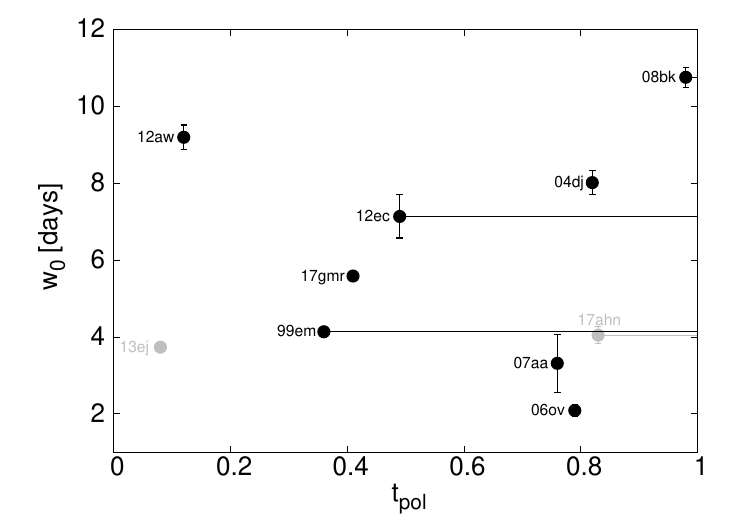}
    \includegraphics[width=\columnwidth]{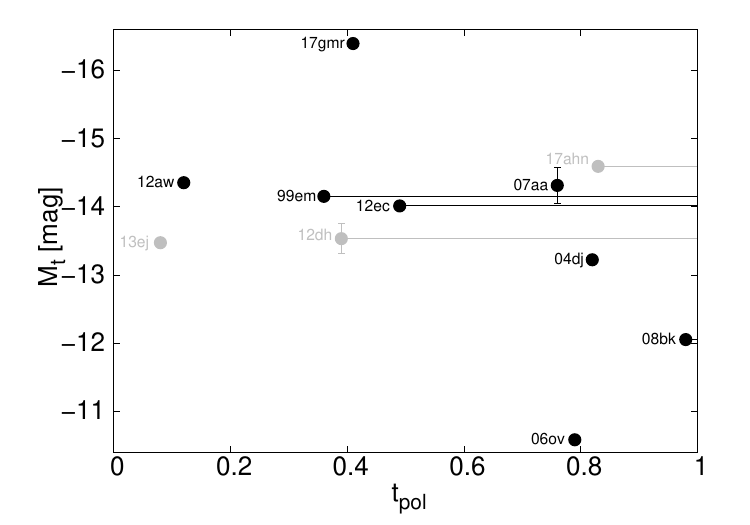}
    \includegraphics[width=\columnwidth]{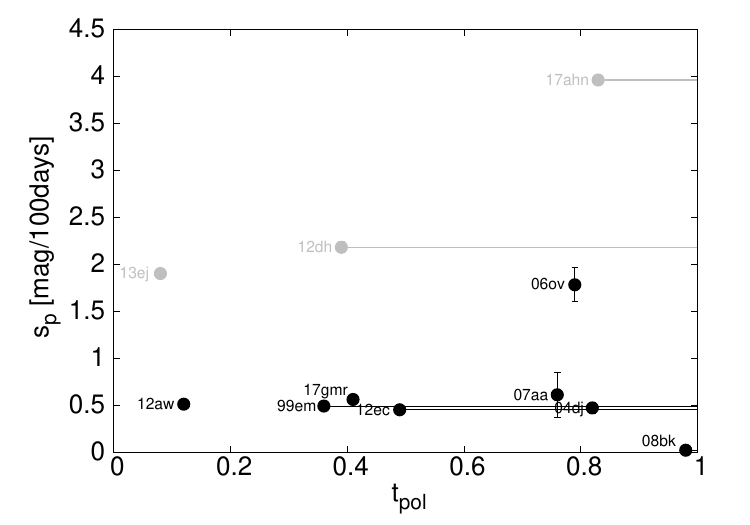}
    \includegraphics[width=\columnwidth]{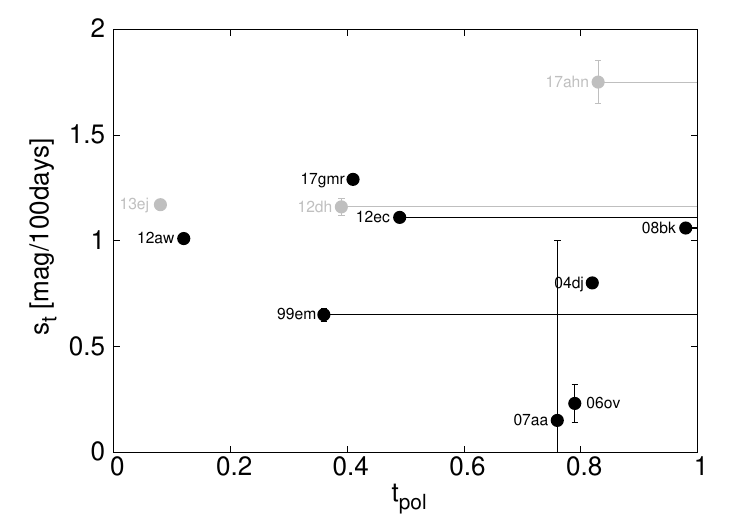}
    \caption{Relations between the polarimetric property ($t_{\rm{pol}}$) and photometric properties ($t_{\rm{\rm{ph}}}$, $a_{0}$, $w_{0}$, $M_{t}$, $s_{\rm{p}}$ and $s_{\rm{t}}$). The black points represent Type~IIP SNe, while gray points do Type~IIL SNe.}
    \label{fig:correlation1}
\end{figure*}

Figure~\ref{fig:correlation1} shows the relations between the polarimetric ($t_{\rm{pol}}$) and photometric parameters ($t_{\rm{\rm{ph}}}$, $a_{0}$, $w_{0}$, $M_{t}$, $s_{\rm{p}}$ and $s_{\rm{t}}$). There seems to be no significant correlation between polarization degree and the other photometric parameters. However, the brightness at the tail phase $M_{t}$ (and thus the $^{56}$Ni mass, see below) and the brightness gap between the photospheric and tail phases ($a_{0}$) might be correlated with $t_{\rm{pol}}$ in Type~IIP SNe, although the limited sample does not allow for a firm conclusion.

\begin{figure*}
    \includegraphics[width=\columnwidth]{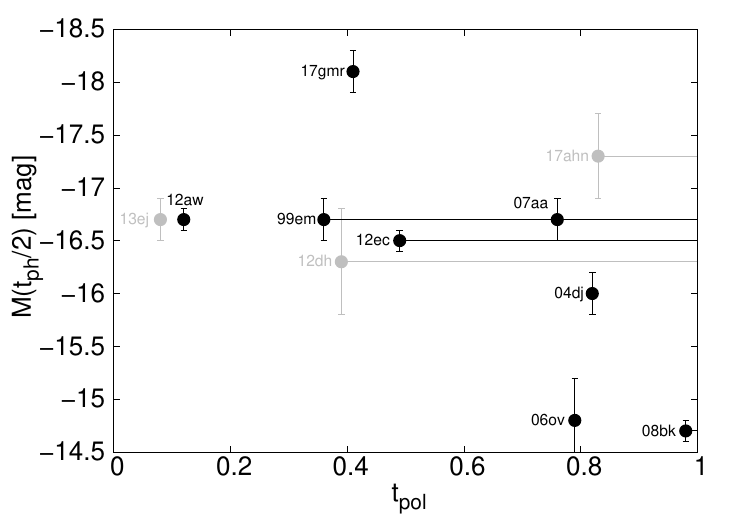}
    \includegraphics[width=\columnwidth]{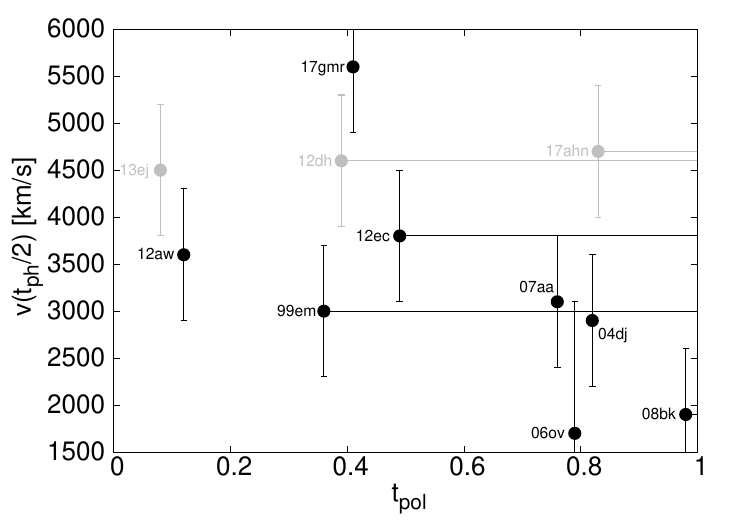}
    \includegraphics[width=\columnwidth]{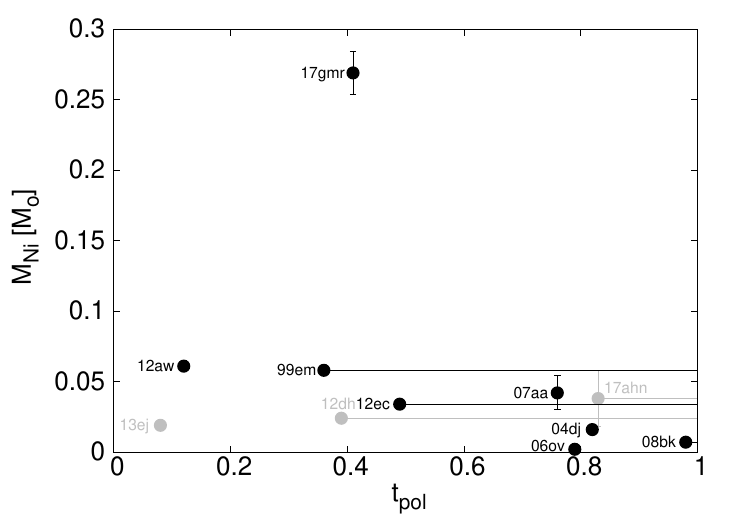}
    \caption{Relations between the polarimetric property ($t_{\rm{pol}}$) and the characteristic values from the photometric and spectroscopic observations ($M(t_{\rm{\rm{ph}}}/2)$, $v(t_{\rm{\rm{ph}}}/2)$ and $M_{\rm{Ni}}$). The black points represent Type~IIP SNe, while gray points do Type~IIL SNe.}
    \label{fig:correlation2}
\end{figure*}

Figure~\ref{fig:correlation2} shows the relations between the polarimetric parameter ($t_{\rm{pol}}$) and the characteristic values from the photometric and spectroscopic observations ($M(t_{\rm{\rm{ph}}}/2)$, $v(t_{\rm{\rm{ph}}}/2)$ and $M_{\rm{Ni}}$). In the Type~IIP sample, there might be some correlations: SNe that are brighter and faster and have larger amount of $^{56}$Ni tend to have more extended aspherical structures. Since the Type~IIL sample is limited, we cannot conclude anything about their behavior. However, they do not seem to follow these correlations. 
We emphasize that the ejecta velocity estimated from the Fe~II absorption minimum is related to the ejecta velocity along the line of sight, and is thus affected by the viewing angle. Most SNe show $\sim 1$ \% of polarization degrees at their polarization peaks, implying aspherical structures with a minimum ellipsoidal axial ratio of 1.2:1.0. If the parts of the ejecta forming the Fe~II line have such an asphericity, the estimated velocities can have an uncertainty of $\pm \gtrsim 10$\% just because of the uncertainties of the viewing angle. This effect is more important for SNe that have more aspherical structures in the line forming regions, i.e. the outer parts of the ejecta above the photosphere.
Moreover, the Fe distribution does not necessarily have to be consistent with the overall aspherical structure. In some extreme cases, e.g., SN~2017gmr, there is a global aspherical structure from the inner core to the outer envelope, which might be created by a jet-like explosion and might produce a more aspherical Fe distribution.
A larger sample is required to statistically confirm this correlation between the ejecta velocity and the polarimetric parameter taking into account this effect.

\begin{figure*}
    \includegraphics[width=2\columnwidth]{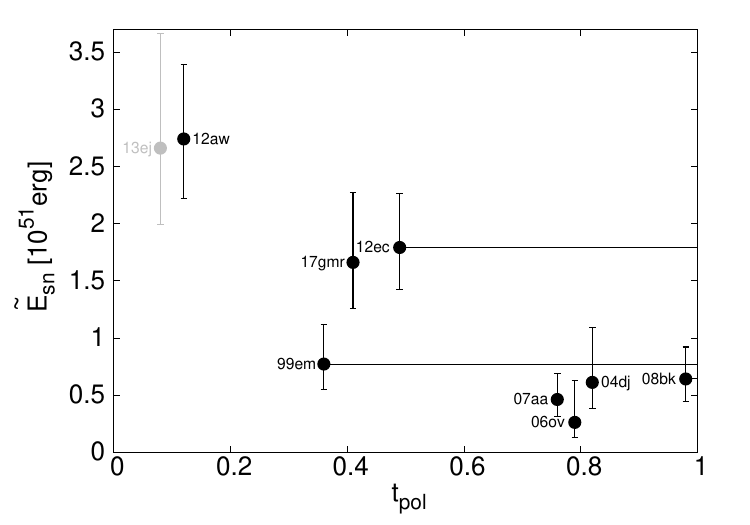}
    \includegraphics[width=\columnwidth]{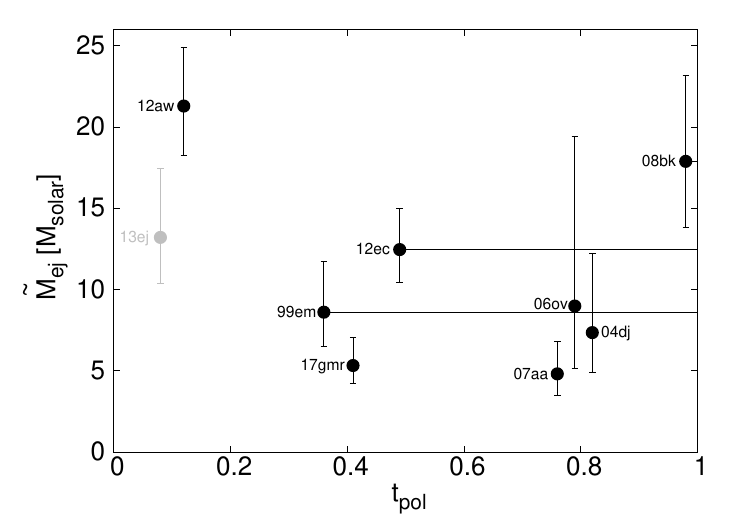}
    \includegraphics[width=\columnwidth]{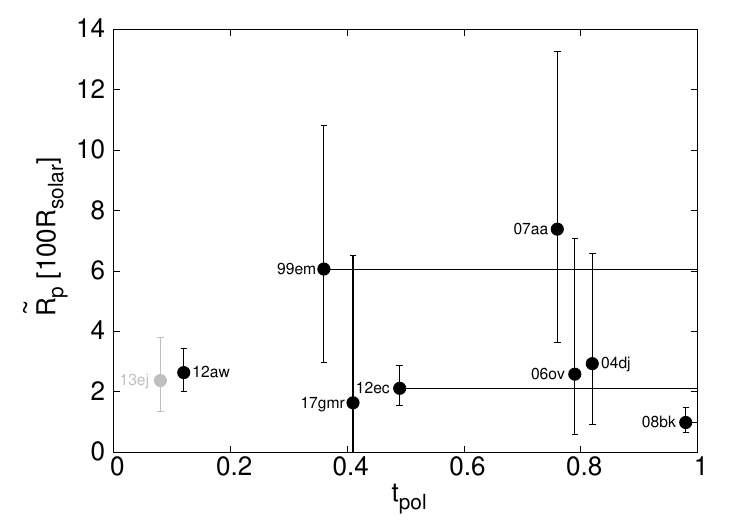}
    \caption{Relations between the polarimetric property ($t_{\rm{pol}}$) and the SN explosion parameters ($\widetilde{M}_{\rm{ej}}$, $\widetilde{E}_{\rm{ej}}$, $\widetilde{R}_{\rm{p}}$). The black points represent Type~IIP SNe, while gray point does the Type~IIL SN~2013ej.}
    \label{fig:correlation3}
\end{figure*}

Figure~\ref{fig:correlation3} shows the relations between the polarimetric property ($t_{\rm{pol}}$) and the SN explosion parameters ($\widetilde{M}_{\rm{ej}}$, $\widetilde{E}_{\rm{ej}}$, $\widetilde{R}_{\rm{p}}$). The polarimetric property ($t_{\rm{pol}}$) is correlated with the explosion energy ($\widetilde{E}_{\rm{ej}}$) rather than the ejecta mass ($\widetilde{M}_{\rm{ej}}$) and the progenitor radius ($\widetilde{R}_{\rm{p}}$): More energetic SNe tend to have more extended aspherical structure in their explosions.
We finally note that the absolute values of the SN explosion parameters are not accurate because of limitation of the one-zone model (see Section~\ref{sec:SN_explosion_properties}). Therefore, only a relative comparison between the various SNe is possible. In addition, the largest uncertainties for the SN explosion parameters come from the uncertainty of the viewing angles which translates into a significant uncertainty on the expansion velocities.

\section{Conclusions} \label{sec:conclusions}

In this article, we have investigated the continuum polarization of 15 Type~II SNe.
Our analysis shows that the objects in the sample have diverse properties. Most SNe show low polarization at early phases with a sudden rise to $\sim 1 \%$ degree at a certain point during the photospheric phase, followed by a slow decline during the tail phase, with a constant polarization angle. The timing of the polarization rise varies for different SNe. 
We interpret this as the signature of different explosion geometries: some SNe have aspherical structures only in their helium cores, while the asphericity of other SNe reaches out to a significant part of the outer hydrogen envelope, with a common axis through the helium core and the hydrogen envelope.
Some SNe show high polarization already at early phases and a change of the polarization angle. This indicates multi-directional aspherical structures in their explosions, which might originate from a combination of a common aspherical SN explosion and an aspherical CSM interaction or from a multi-directional SN explosion.
An exception in our sample is the Type~IIL SN~2017ahn, which has low polarization during both the photospheric and tail phases. Since this object is unique in our sample, we cannot conclude whether this SN is produced by a completely spherical explosion or just the result of a different viewing angle for a similarly aspherical explosion.

The timing of the polarization rise in Type~IIP SNe appears to be correlated with their brightness, ejecta velocity, and Ni mass: SNe with higher brightness, higher ejecta velocity and larger Ni production show more extended asphericities. In particular, there is a clear correlation between the timing of the polarization rise and the explosion energy: the asphericity in the explosions is proportional to the explosion energy (see Section~\ref{sec:correlation} and Figure~\ref{fig:correlation3}). This implies that the emergence of a global aspherical structure, e.g., a jet-like structure, might be the key ingredient in the explosion mechanism to produce an energetic SN.

\begin{acknowledgements}
    We thank Keiichi Maeda, Masaomi Tanaka and Giuliano Pignata for useful discussions.
    T.N.\ was supported by a Japan Society for the Promotion of Science (JSPS) Overseas Research Fellowship and was funded by the Academy of Finland project 328898.
    H.K.\ was funded by the Academy of Finland projects 324504 and 328898.
    S.M. was funded by the Academy of Finland project 350458.
    This research has made use of NASA's Astrophysics Data System Bibliographic Services.
\end{acknowledgements}


\bibliographystyle{aa} % style aa.bst
\bibliography{aa.bib}

%%%%%%%%%%%%%%%%%%%%%%%%%%%%%%%%%%%%%%%%%%%%%%%%%%

%%%%%%%%%%%%%%%%% APPENDICES %%%%%%%%%%%%%%%%%%%%%

%\newpage

\appendix

\section{Polarization degrees and angles of the continuum polarization for the VLT sample} \label{sec:app_A}

\begin{table*}
      \caption[]{SN 2017gmr}
      \label{tab:log_17gmr}
\scalebox{0.80}[0.80]{$\displaystyle
         \begin{array}{lccccc}
            \hline
            \noalign{\smallskip}
            \rm{Date} & \rm{MJD} & \rm{Phase}^{a} & \rm{Days \;from \;explosion} & \rm{Pol.\;degree} & \rm{Pol. \;angle}\\
            (\rm{UT}) & (\rm{days}) & (\rm{days}) & (\rm{days}) & (\rm{per\;cent}) & (\rm{degrees})\\
            \noalign{\smallskip}
            \hline\hline
            \noalign{\smallskip}      
            2017-10-19.60 & 58045.60 & -46.41 & +46.51 & 0.18 \pm 0.01 & 87.5 \pm 2.0\\
            \noalign{\smallskip} \hline \noalign{\smallskip}
            2017-11-10.24 & 58067.24 & -24.77 & +68.15 & 0.71 \pm 0.02 & 93.4 \pm 0.7 \\
            \noalign{\smallskip} \hline \noalign{\smallskip}
            2017-12-12.86 & 58099.86 & +7.85 & +100.77 & 0.81 \pm 0.01 & 99.7 \pm 0.7 \\
            \noalign{\smallskip} \hline \noalign{\smallskip}
            2017-12-21.76 & 58108.76 & +16.75 & +109.67 & 0.59 \pm 0.01 & 97.4 \pm 0.7 \\
            \noalign{\smallskip} \hline \noalign{\smallskip}
            2018-01-16.93 & 58134.93 & +42.92 & +135.84 & 0.39 \pm 0.02 & 93.9 \pm 2.1 \\
            \noalign{\smallskip}
            \hline
         \end{array}
         $}
         
         \begin{minipage}{\hsize}
        \smallskip
        Notes. %${}^{a}$Relative to $t_{0}=58100$ (MJD), which is the timing of the end of the photospheric phase. ${}^{b}$Relative to $t=58000.20$ (MJD), which is the timing of the first detection and almost the explosion time. The non-detection was reported 1.97 days before the first detection. For the sake of increasing the signal-to-noise ratio, we combine the spectropolarimetric data into five groups, where we have checked the consistency of the spectra within each group.
        \end{minipage}
\end{table*}

\begin{table*}
      \caption[]{SN 2017ahn}
      \label{tab:log_17ahn}
\scalebox{0.80}[0.80]{$\displaystyle
         \begin{array}{lccccc}
            \hline
            \noalign{\smallskip}
            \rm{Date} & \rm{MJD} & \rm{Phase}^{a} & \rm{Days \;from \;explosion} & \rm{Pol.\;degree} & \rm{Pol. \;angle}\\
            (\rm{UT}) & (\rm{days}) & (\rm{days}) & (\rm{days}) & (\rm{per\;cent}) & (\rm{degrees})\\
            \noalign{\smallskip}
            \hline\hline
            \noalign{\smallskip}      
            2017-03-24.75 & 57836.75 & -16.63 & +44.99 & 0.13 \pm 0.02 & 11.0 \pm 3.4\\
            \noalign{\smallskip} \hline \noalign{\smallskip}
            2017-03-31.38 & 57843.38 & -10.00 & +51.62 & 0.09 \pm 0.03 & 28.2 \pm 4.3\\
            \noalign{\smallskip} \hline \noalign{\smallskip}
            2017-04-24.06 & 57867.06 & +13.68 & +75.30 & 0.16 \pm 0.07 & 29.4 \pm 6.0\\
            \noalign{\smallskip} \hline \noalign{\smallskip}
            2017-04-28.73 & 57871.73 & +18.35 & +79.97 & 0.15 \pm 0.08 & 39.2 \pm 6.8\\
            \noalign{\smallskip}
            \hline
         \end{array}
         $}
         
         \begin{minipage}{\hsize}
        \smallskip
        Notes. %${}^{a}$Relative to $t_{0}=58100$ (MJD), which is the timing of the end of the photospheric phase. ${}^{b}$Relative to $t=58000.20$ (MJD), which is the timing of the first detection and almost the explosion time. The non-detection was reported 1.97 days before the first detection. For the sake of increasing the signal-to-noise ratio, we combine the spectropolarimetric data into five groups, where we have checked the consistency of the spectra within each group.
        \end{minipage}
\end{table*}

\begin{table*}
      \caption[]{SN 2013ej}
      \label{tab:log_13ej}
\scalebox{0.80}[0.80]{$\displaystyle
         \begin{array}{lcccccc}
            \hline
            \noalign{\smallskip}
            \rm{Date} & \rm{MJD} & \rm{Phase}^{a} & \rm{Days \;from \;explosion} & \rm{Pol.\;degree} & \rm{Pol. \;angle}\\
            (\rm{UT}) & (\rm{days}) & (\rm{days}) & (\rm{days}) & (\rm{per\;cent}) & (\rm{degrees})\\
            \noalign{\smallskip}
            \hline\hline
            \noalign{\smallskip}      
            2013-08-01.38 & 56505.38 & -91.99 & +8.48 & 0.40 (0.00) \pm 0.01 & 78.4 (-) \pm 0.6 \\
            \noalign{\smallskip} \hline \noalign{\smallskip}
            2013-08-27.28 & 56531.28 & -66.09 & +34.38 & 0.75 (0.37) \pm 0.01 & 84.0 (90.0) \pm 0.3 \\
            \noalign{\smallskip} \hline \noalign{\smallskip}
            2013-09-17.21 & 56552.21 & -45.16 & +55.31 & 1.03 (0.67) \pm 0.01 & 88.9 (94.9) \pm 0.2 \\
            \noalign{\smallskip} \hline \noalign{\smallskip}
            2013-09-29.24 & 56564.24 & -33.13 & +67.34 & 1.24 (0.90) \pm 0.01 & 92.1 (97.9) \pm 0.2 \\
            \noalign{\smallskip} \hline \noalign{\smallskip}
            2013-10-29.23 & 56594.23 & -3.14 & +97.33 & 1.19 \pm 0.01 & 97.2  \pm 0.3 \\
            \noalign{\smallskip} \hline \noalign{\smallskip}
            2013-12-04.56 & 56630.56 & +33.19 & +133.66 & 0.54 \pm 0.02 & 106.1 \pm 1.5 \\ 
            \noalign{\smallskip} \hline \noalign{\smallskip}
            2014-01-09.38 & 56666.38 & +69.01 & +169.48 & 0.64 \pm 0.05 & 123.5 \pm 3.6 \\
            \noalign{\smallskip}
            \hline
         \end{array}
         $}
         
         \begin{minipage}{\hsize}
        \smallskip
        Notes. %${}^{a}$Relative to $t_{0}=58100$ (MJD), which is the timing of the end of the photospheric phase. ${}^{b}$Relative to $t=58000.20$ (MJD), which is the timing of the first detection and almost the explosion time. The non-detection was reported 1.97 days before the first detection. For the sake of increasing the signal-to-noise ratio, we combine the spectropolarimetric data into five groups, where we have checked the consistency of the spectra within each group.
        \end{minipage}
\end{table*}

\begin{table*}
      \caption[]{SN 2012ec}
      \label{tab:log_12ec}
\scalebox{0.80}[0.80]{$\displaystyle
         \begin{array}{lcccccc}
            \hline
            \noalign{\smallskip}
            \rm{Date} & \rm{MJD} & \rm{Phase}^{a} & \rm{Days \;from \;explosion} & \rm{Pol.\;degree} & \rm{Pol. \;angle}\\
            (\rm{UT}) & (\rm{days}) & (\rm{days}) & (\rm{days}) & (\rm{per\;cent}) & (\rm{degrees})\\
            \noalign{\smallskip}
            \hline\hline
            \noalign{\smallskip}      
            2012-09-08.28 & 56178.28 & -71.42 & +35.28 & 0.05 \pm 0.02 & 131.5 \pm 5.9 \\
            \noalign{\smallskip} \hline \noalign{\smallskip}
            2012-09-25.21 & 56195.21 & -54.49 & +52.21 & 0.12 \pm 0.02 & 134.5 \pm 5.1 \\
            \noalign{\smallskip}
            \hline
         \end{array}
         $}
         
         \begin{minipage}{\hsize}
        \smallskip
        Notes. %${}^{a}$Relative to $t_{0}=58100$ (MJD), which is the timing of the end of the photospheric phase. ${}^{b}$Relative to $t=58000.20$ (MJD), which is the timing of the first detection and almost the explosion time. The non-detection was reported 1.97 days before the first detection. For the sake of increasing the signal-to-noise ratio, we combine the spectropolarimetric data into five groups, where we have checked the consistency of the spectra within each group.
        \end{minipage}
\end{table*}

\begin{table*}
      \caption[]{SN 2012dh}
      \label{tab:log_12dh}
\scalebox{0.80}[0.80]{$\displaystyle
         \begin{array}{lcccccc}
            \hline
            \noalign{\smallskip}
            \rm{Date} & \rm{MJD} & \rm{Phase}^{a} & \rm{Days \;from \;explosion} & \rm{Pol.\;degree} & \rm{Pol. \;angle}\\
            (\rm{UT}) & (\rm{days}) & (\rm{days}) & (\rm{days}) & (\rm{per\;cent}) & (\rm{degrees})\\
            \noalign{\smallskip}
            \hline\hline
            \noalign{\smallskip}      
            2012-07-18.03 & 56126.03 & -94.35 \pm 16.40 & +21.05 & 0.05 \pm 0.01 & 76.3 \pm 3.9 \\
            \noalign{\smallskip} \hline \noalign{\smallskip}
            2012-08-11.98 & 56150.98 & -69.40 \pm 16.40 & +46.00 & 0.25 \pm 0.02 & 21.5 \pm 2.2 \\
            \noalign{\smallskip} \hline \noalign{\smallskip}
            2012-08-24.99 & 56163.99 & -56.39 \pm 16.40 & +59.01 & 0.12 \pm 0.03 & 173.0 \pm 6.7 \\
            \noalign{\smallskip} \hline \noalign{\smallskip}
            2012-08-27.01 & 56166.01 & -54.37 \pm 16.40 & +61.03 & 0.19 \pm 0.05 & 173.7 \pm 5.2 \\
            \noalign{\smallskip}
            \hline
         \end{array}
         $}
         
         \begin{minipage}{\hsize}
        \smallskip
        Notes. %${}^{a}$Relative to $t_{0}=58100$ (MJD), which is the timing of the end of the photospheric phase. ${}^{b}$Relative to $t=58000.20$ (MJD), which is the timing of the first detection and almost the explosion time. The non-detection was reported 1.97 days before the first detection. For the sake of increasing the signal-to-noise ratio, we combine the spectropolarimetric data into five groups, where we have checked the consistency of the spectra within each group.
        \end{minipage}
\end{table*}

\begin{table*}
      \caption[]{SN 2012aw}
      \label{tab:log_12aw}
\scalebox{0.80}[0.80]{$\displaystyle
         \begin{array}{lcccccc}
            \hline
            \noalign{\smallskip}
            \rm{Date} & \rm{MJD} & \rm{Phase}^{a} & \rm{Days \;from \;explosion} & \rm{Pol.\;degree} & \rm{Pol. \;angle}\\
            (\rm{UT}) & (\rm{days}) & (\rm{days}) & (\rm{days}) & (\rm{per\;cent}) & (\rm{degrees})\\
            \noalign{\smallskip}
            \hline\hline
            \noalign{\smallskip}      
            2012-04-01.19 & 56018.19 & -117.74 & +16.09 & 0.35 \pm 0.01 & 34.8 \pm 1.5 \\
            \noalign{\smallskip} \hline \noalign{\smallskip}
            2012-05-01.04 & 56048.04 & -87.89 & +45.94 & 0.31 \pm 0.02 & 22.8 \pm 2.6 \\
            \noalign{\smallskip} \hline \noalign{\smallskip}
            2012-05-18.07 & 56065.07 & -70.86 & +62.97 & 0.15 \pm 0.01 & 86.3 \pm 1.6 \\
            \noalign{\smallskip} \hline \noalign{\smallskip}
            2012-05-27.01 & 56074.01 & -61.92 & +71.91 & 0.16 \pm 0.01 & 102.6 \pm 3.3 \\
            \noalign{\smallskip} \hline \noalign{\smallskip}
            2012-06-16.03 & 56094.03 & -41.90 & +91.93 & 0.45 \pm 0.01 & 117.8  \pm 0.5 \\
            \noalign{\smallskip} \hline \noalign{\smallskip}
            2012-07-02.01 & 56110.01 & -25.92 & +107.91 & 0.75 \pm 0.01 & 124.1 \pm 0.4 \\ 
            \noalign{\smallskip} \hline \noalign{\smallskip}
            2012-07-15.98 & 56123.98 & -11.95 & +121.88 & 1.00 \pm 0.03 & 126.3 \pm 1.4 \\
            \noalign{\smallskip}
            \hline
         \end{array}
         $}
         
         \begin{minipage}{\hsize}
        \smallskip
        Notes. %${}^{a}$Relative to $t_{0}=58100$ (MJD), which is the timing of the end of the photospheric phase. ${}^{b}$Relative to $t=58000.20$ (MJD), which is the timing of the first detection and almost the explosion time. The non-detection was reported 1.97 days before the first detection. For the sake of increasing the signal-to-noise ratio, we combine the spectropolarimetric data into five groups, where we have checked the consistency of the spectra within each group.
        \end{minipage}
\end{table*}

\begin{table*}
      \caption[]{SN 2010hv}
      \label{tab:log_10hv}
\scalebox{0.80}[0.80]{$\displaystyle
         \begin{array}{lcccccc}
            \hline
            \noalign{\smallskip}
            \rm{Date} & \rm{MJD} & \rm{Phase}^{a} & \rm{Days \;from \;explosion} & \rm{Pol.\;degree} & \rm{Pol. \;angle}\\
            (\rm{UT}) & (\rm{days}) & (\rm{days}) & (\rm{days}) & (\rm{per\;cent}) & (\rm{degrees})\\
            \noalign{\smallskip}
            \hline\hline
            \noalign{\smallskip}      
            2010-09-21.15 & 55460.15 & -73.45 \pm 24.4 & +41.95 \pm 8.00 & 0.06 \pm 0.05 & 89.5 \pm 4.6 \\
            \noalign{\smallskip} \hline \noalign{\smallskip}
            2010-09-24.14 & 55463.14 & -70.46 \pm 24.4 & +44.94 \pm 8.00 & 0.11 \pm 0.05 & 120.6 \pm 4.1 \\
            \noalign{\smallskip} \hline \noalign{\smallskip}
            2010-10-15.07 & 55484.07 & -49.53 \pm 24.4 & +65.87 \pm 8.00 & 0.06 \pm 0.04 & 40.8 \pm 4.4 \\
            \noalign{\smallskip} \hline \noalign{\smallskip}
            2010-11-29.56 & 55529.56 & -4.04 \pm 24.4 & +111.36 \pm 8.00 & 0.11 \pm 0.02 & 18.1 \pm 3.7 \\
            \noalign{\smallskip}
            \hline
         \end{array}
         $}
         
         \begin{minipage}{\hsize}
        \smallskip
        Notes. %${}^{a}$Relative to $t_{0}=58100$ (MJD), which is the timing of the end of the photospheric phase. ${}^{b}$Relative to $t=58000.20$ (MJD), which is the timing of the first detection and almost the explosion time. The non-detection was reported 1.97 days before the first detection. For the sake of increasing the signal-to-noise ratio, we combine the spectropolarimetric data into five groups, where we have checked the consistency of the spectra within each group.
        \end{minipage}
\end{table*}

\begin{table*}
      \caption[]{SN 2010co}
      \label{tab:log_10co}
\scalebox{0.80}[0.80]{$\displaystyle
         \begin{array}{lcccccc}
            \hline
            \noalign{\smallskip}
            \rm{Date} & \rm{MJD} & \rm{Phase}^{a} & \rm{Days \;from \;explosion} & \rm{Pol.\;degree} & \rm{Pol. \;angle}\\
            (\rm{UT}) & (\rm{days}) & (\rm{days}) & (\rm{days}) & (\rm{per\;cent}) & (\rm{degrees})\\
            \noalign{\smallskip}
            \hline\hline
            \noalign{\smallskip}      
            2010-06-05.73 & 55352.73 & -73.27 \pm 27.90 & +42.13 \pm 11.50 & 0.21 \pm 0.06 & 115.1 \pm 4.2 \\
            \noalign{\smallskip} \hline \noalign{\smallskip}
            2010-07-08.21 & 55385.21 & -40.79 \pm 27.90 & +74.61 \pm 11.50 & 0.83 \pm 0.10 & 175.2 \pm 5.2 \\
            \noalign{\smallskip}
            \hline
         \end{array}
         $}
         
         \begin{minipage}{\hsize}
        \smallskip
        Notes. %${}^{a}$Relative to $t_{0}=58100$ (MJD), which is the timing of the end of the photospheric phase. ${}^{b}$Relative to $t=58000.20$ (MJD), which is the timing of the first detection and almost the explosion time. The non-detection was reported 1.97 days before the first detection. For the sake of increasing the signal-to-noise ratio, we combine the spectropolarimetric data into five groups, where we have checked the consistency of the spectra within each group.
        \end{minipage}
\end{table*}

\begin{table*}
      \caption[]{SN 2008bk}
      \label{tab:log_08bk}
\scalebox{0.80}[0.80]{$\displaystyle
         \begin{array}{lcccccc}
            \hline
            \noalign{\smallskip}
            \rm{Date} & \rm{MJD} & \rm{Phase}^{a} & \rm{Days \;from \;explosion} & \rm{Pol.\;degree} & \rm{Pol. \;angle}\\
            (\rm{UT}) & (\rm{days}) & (\rm{days}) & (\rm{days}) & (\rm{per\;cent}) & (\rm{degrees})\\
            \noalign{\smallskip}
            \hline\hline
            \noalign{\smallskip}      
            2008-06-02.36 & 54619.36 & -60.24 & +69.86 & 0.19 \pm 0.01 & 148.3 \pm 2.5 \\
            \noalign{\smallskip} \hline \noalign{\smallskip}
            2008-07-01.31 & 54648.31 & -31.29 & +98.81 & 0.44 \pm 0.01 & 142.2 \pm 0.9 \\
            \noalign{\smallskip} \hline \noalign{\smallskip}
            2008-07-24.29 & 54671.29 & -8.31 & +121.79 & 0.23 \pm 0.01 & 133.6 \pm 0.8 \\
            \noalign{\smallskip} \hline \noalign{\smallskip}
            2008-07-30.28 & 54677.28 & -2.32 & +127.78 & 0.31 \pm 0.01 & 117.6 \pm 2.0 \\
            \noalign{\smallskip} \hline \noalign{\smallskip}
            2008-08-06.25 & 54684.25 & +4.65 & +134.75 & 0.43 \pm 0.01 & 102.3 \pm 1.0 \\
            \noalign{\smallskip} \hline \noalign{\smallskip}
            2008-09-28.10 & 54737.10 & +57.50 & +187.60 & 1.17 \pm 0.01 & 126.3 \pm 0.4 \\
            \noalign{\smallskip} \hline \noalign{\smallskip}
            2008-11-19.03 & 54789.03 & +109.43 & +239.53 & 0.35 \pm 0.02 & 112.7 \pm 1.9 \\
            \noalign{\smallskip} \hline \noalign{\smallskip}
            2008-12-20.04 & 54820.04 & +140.44 & +270.54 & 0.32 \pm 0.02 & 108.2 \pm 2.5 \\
            \noalign{\smallskip} \hline \noalign{\smallskip}
            2009-01-01.04 & 54832.04 & +152.44 & +282.54 & 0.32 \pm 0.03 & 110.9 \pm 2.7 \\
            \noalign{\smallskip}
            \hline
         \end{array}
         $}
         
         \begin{minipage}{\hsize}
        \smallskip
        Notes. %${}^{a}$Relative to $t_{0}=58100$ (MJD), which is the timing of the end of the photospheric phase. ${}^{b}$Relative to $t=58000.20$ (MJD), which is the timing of the first detection and almost the explosion time. The non-detection was reported 1.97 days before the first detection. For the sake of increasing the signal-to-noise ratio, we combine the spectropolarimetric data into five groups, where we have checked the consistency of the spectra within each group.
        \end{minipage}
\end{table*}

\begin{table*}
      \caption[]{SN 2001du}
      \label{tab:log_01du}
\scalebox{0.80}[0.80]{$\displaystyle
         \begin{array}{lcccccc}
            \hline
            \noalign{\smallskip}
            \rm{Date} & \rm{MJD} & \rm{Phase}^{a} & \rm{Days \;from \;explosion} & \rm{Pol.\;degree} & \rm{Pol. \;angle}\\
            (\rm{UT}) & (\rm{days}) & (\rm{days}) & (\rm{days}) & (\rm{per\;cent}) & (\rm{degrees})\\
            \noalign{\smallskip}
            \hline\hline
            \noalign{\smallskip}      
            2001-08-30.32 & 52151.32 & -109.28 \pm 16.90 & +6.12 \pm 0.50 & 0.05 \pm 0.04 & 143.6 \pm 10.2 \\
            \noalign{\smallskip} \hline \noalign{\smallskip}
            2001-09-13.36 & 52165.36 & -95.24 \pm 16.90 & +20.16 \pm 0.50 & 0.13 \pm 0.04 & 3.7 \pm 8.5 \\
            \noalign{\smallskip}
            \hline
         \end{array}
         $}
         
         \begin{minipage}{\hsize}
        \smallskip
        Notes. %${}^{a}$Relative to $t_{0}=58100$ (MJD), which is the timing of the end of the photospheric phase. ${}^{b}$Relative to $t=58000.20$ (MJD), which is the timing of the first detection and almost the explosion time. The non-detection was reported 1.97 days before the first detection. For the sake of increasing the signal-to-noise ratio, we combine the spectropolarimetric data into five groups, where we have checked the consistency of the spectra within each group.
        \end{minipage}
\end{table*}

\begin{table*}
      \caption[]{SN 2001dh}
      \label{tab:log_01dh}
\scalebox{0.80}[0.80]{$\displaystyle
         \begin{array}{lcccccc}
            \hline
            \noalign{\smallskip}
            \rm{Date} & \rm{MJD} & \rm{Phase}^{a} & \rm{Days \;from \;explosion} & \rm{Pol.\;degree} & \rm{Pol. \;angle}\\
            (\rm{UT}) & (\rm{days}) & (\rm{days}) & (\rm{days}) & (\rm{per\;cent}) & (\rm{degrees})\\
            \noalign{\smallskip}
            \hline\hline
            \noalign{\smallskip}      
            2001-08-10.18 & 52131.18 & -83.72 \pm 29.80 & +31.68 \pm 13.40 & 0.97 \pm 0.05 & 154.6 \pm 1.9 \\
            \noalign{\smallskip} \hline \noalign{\smallskip}
            2001-08-19.16 & 52140.16 & -74.74 \pm 29.80 & +40.66 \pm 13.40 & 0.89 \pm 0.05 & 151.8 \pm 2.3 \\
            \noalign{\smallskip} \hline \noalign{\smallskip}
            2001-08-29.05 & 52150.05 & -64.85 \pm 29.80 & +50.55 \pm 13.40 & 0.74 \pm 0.07 & 154.7 \pm 4.0 \\
            \noalign{\smallskip} \hline \noalign{\smallskip}
            2001-09-11.09 & 52163.09 & -51.81 \pm 29.80 & +63.59 \pm 13.40 & 0.70 \pm 0.14 & 162.2 \pm 8.0 \\
            \noalign{\smallskip}
            \hline
         \end{array}
         $}
         
         \begin{minipage}{\hsize}
        \smallskip
        Notes. %${}^{a}$Relative to $t_{0}=58100$ (MJD), which is the timing of the end of the photospheric phase. ${}^{b}$Relative to $t=58000.20$ (MJD), which is the timing of the first detection and almost the explosion time. The non-detection was reported 1.97 days before the first detection. For the sake of increasing the signal-to-noise ratio, we combine the spectropolarimetric data into five groups, where we have checked the consistency of the spectra within each group.
        \end{minipage}
\end{table*}

\section{Spectropolarimetric data of the FORS/VLT sample} \label{sec:app_B}

\begin{figure*}
    \includegraphics[width=2.0\columnwidth]{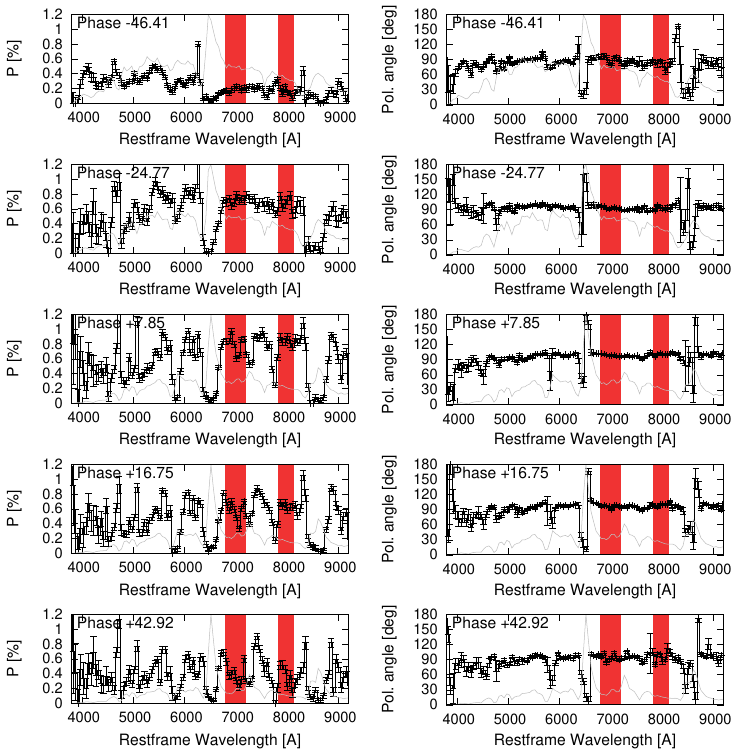}
    \caption{Spectropolarimetry of SN 2017gmr after ISP subtraction. The red hatching shows the adopted wavelength range for the continuum polarization estimate.}
    \label{fig:app_17gmr}
\end{figure*}

\begin{figure*}
    \includegraphics[width=2.0\columnwidth]{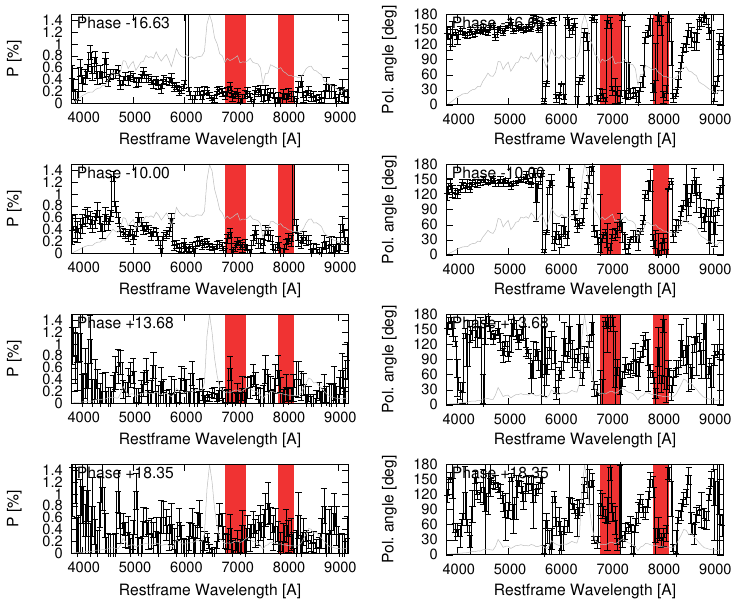}
    \caption{Same as Figure~\ref{fig:app_17gmr}, but for SN 2017ahn.}
    \label{fig:app_17ahn}
\end{figure*}

\begin{figure*}
    \includegraphics[width=2.0\columnwidth]{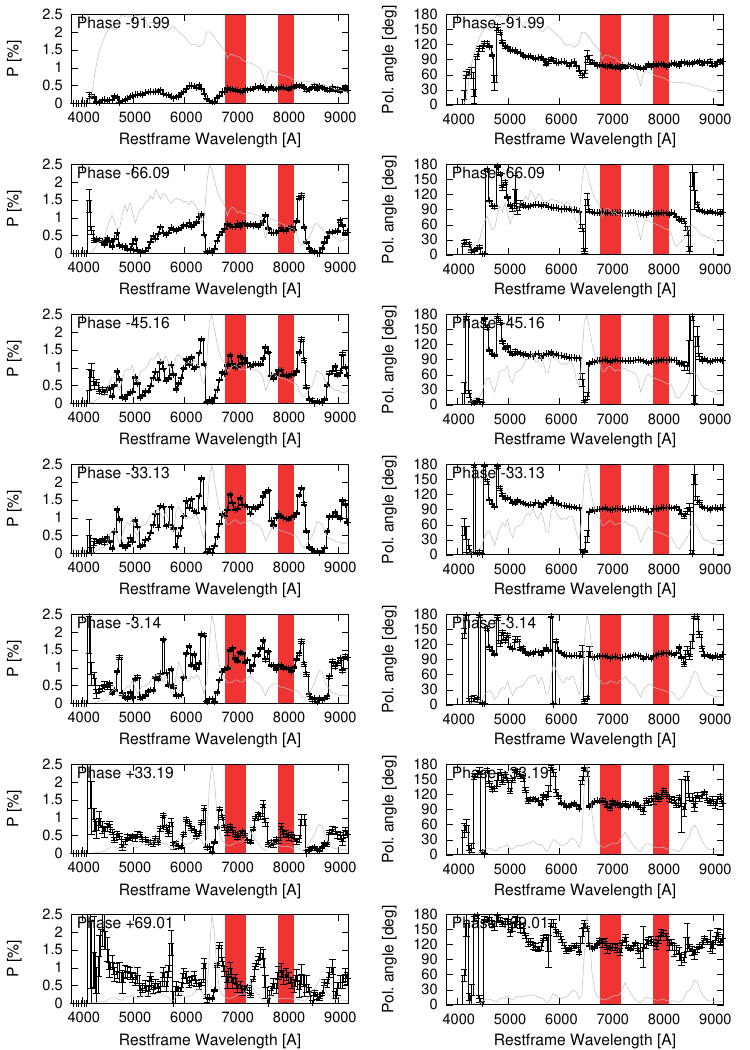}
    \caption{Same as Figure~\ref{fig:app_17gmr}, but for SN 2013ej.}
    \label{fig:app_13ej}
\end{figure*}

\begin{figure*}
    \includegraphics[width=2.0\columnwidth]{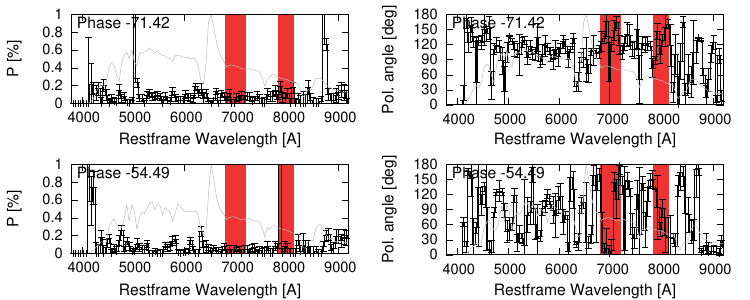}
    \caption{Same as Figure~\ref{fig:app_17gmr}, but for SN 2012ec.}
    \label{fig:app_12ec}
\end{figure*}

\begin{figure*}
    \includegraphics[width=2.0\columnwidth]{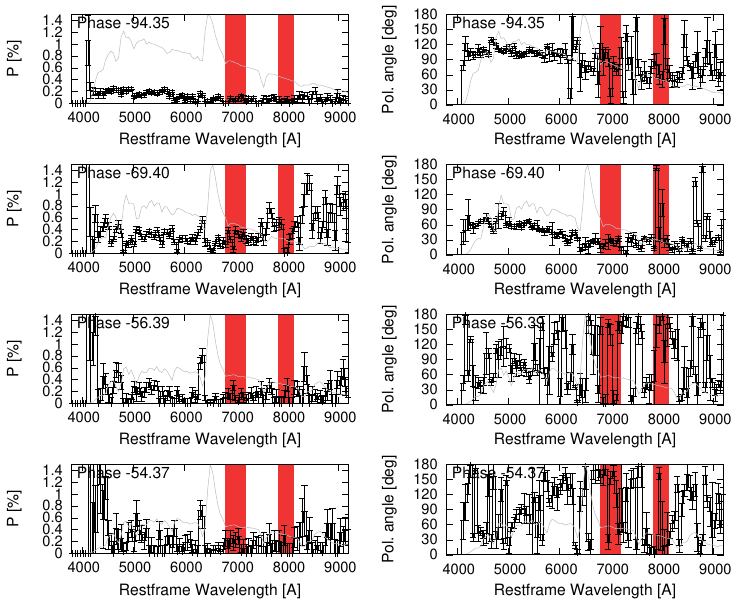}
    \caption{Same as Figure~\ref{fig:app_17gmr}, but for SN 2012dh.}
    \label{fig:app_12dh}
\end{figure*}

\begin{figure*}
    \includegraphics[width=2.0\columnwidth]{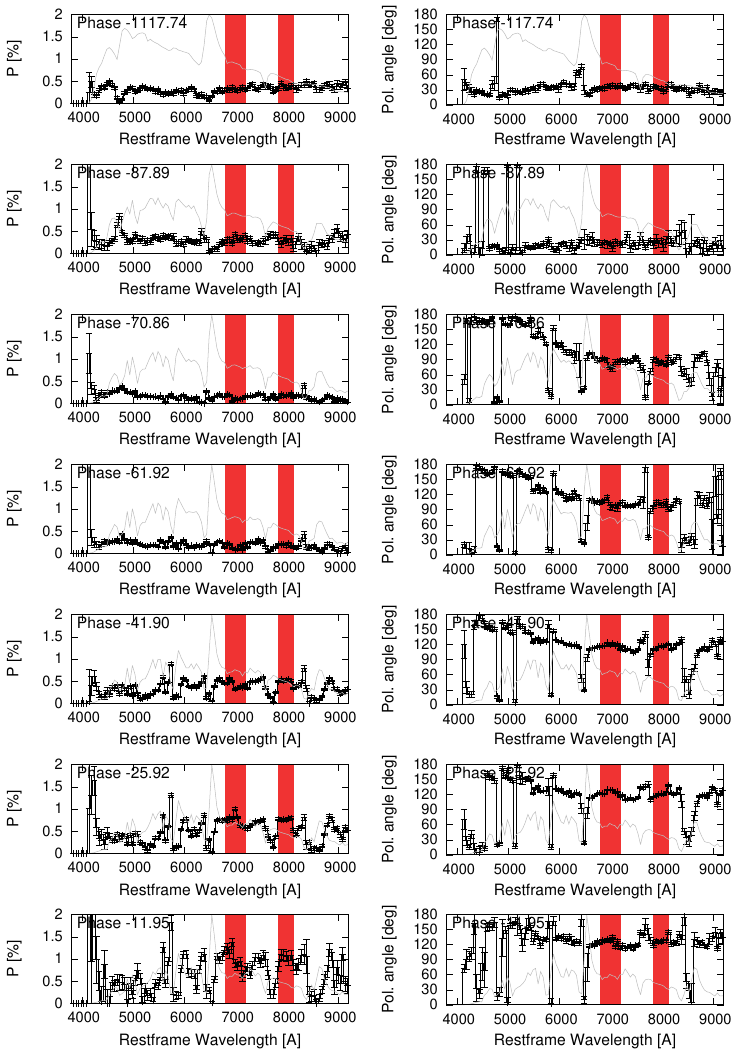}
    \caption{Same as Figure~\ref{fig:app_17gmr}, but for SN 2012aw.}
    \label{fig:app_12aw}
\end{figure*}

\begin{figure*}
    \includegraphics[width=2.0\columnwidth]{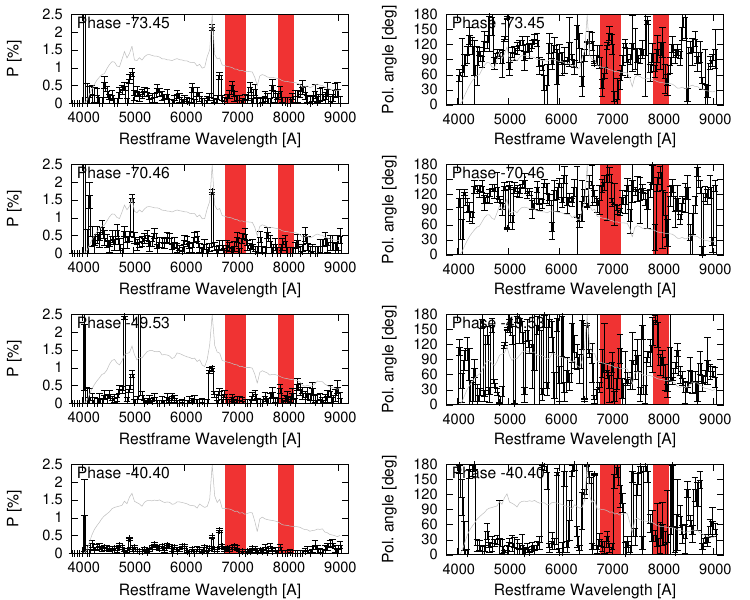}
    \caption{Same as Figure~\ref{fig:app_17gmr}, but for SN 2010hv.}
    \label{fig:app_10hv}
\end{figure*}

\begin{figure*}
    \includegraphics[width=2.0\columnwidth]{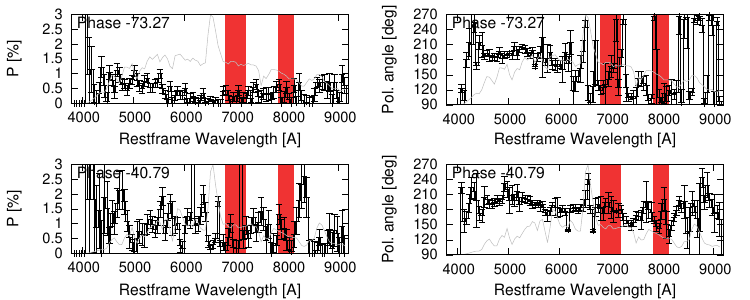}
    \caption{Same as Figure~\ref{fig:app_17gmr}, but for SN 2010co.}
    \label{fig:app_10co}
\end{figure*}

\begin{figure*}
    \includegraphics[width=1.5\columnwidth]{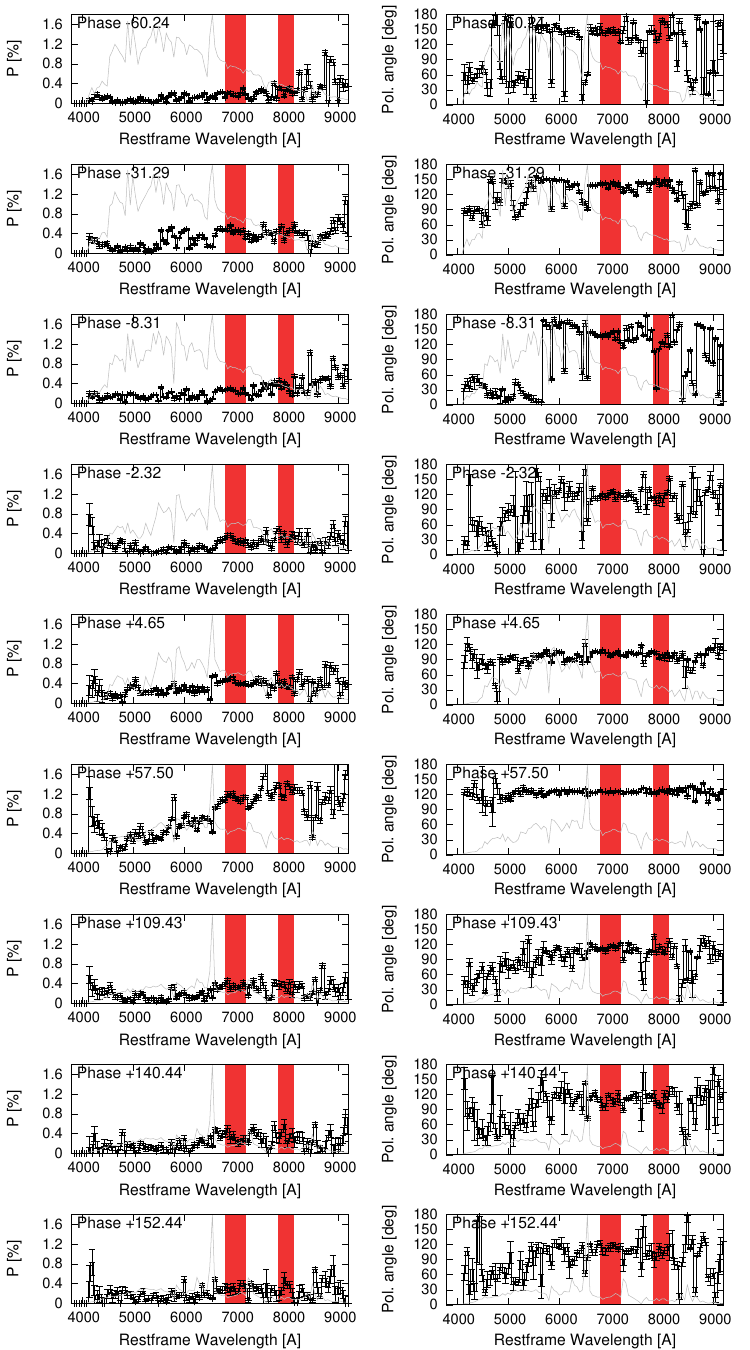}
    \caption{Same as Figure~\ref{fig:app_17gmr}, but for SN 2008bk.}
    \label{fig:app_08bk}
\end{figure*}

\begin{figure*}
    \includegraphics[width=2.0\columnwidth]{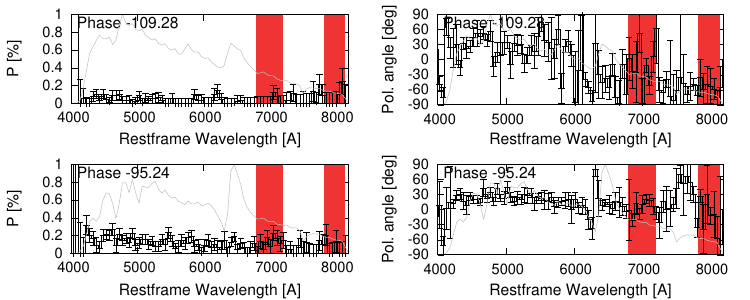}
    \caption{Same as Figure~\ref{fig:app_17gmr}, but for SN 2001du.}
    \label{fig:app_01du}
\end{figure*}

\begin{figure*}
    \includegraphics[width=2.0\columnwidth]{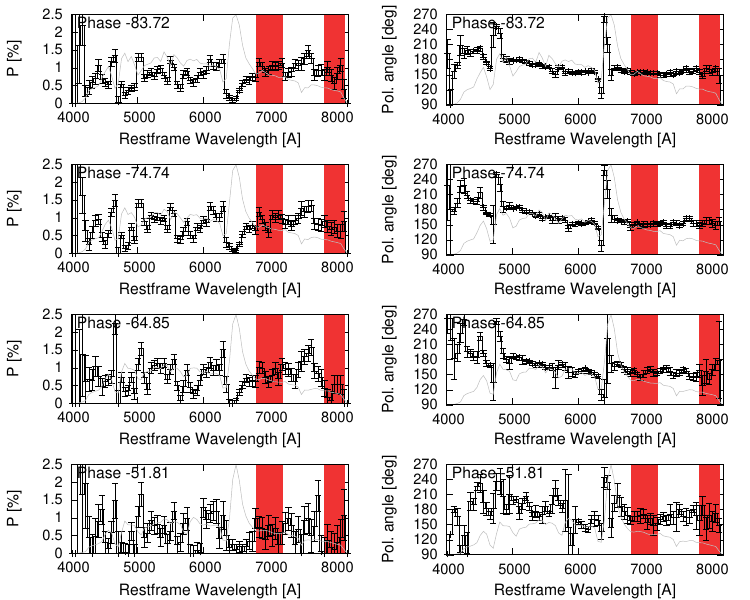}
    \caption{Same as Figure~\ref{fig:app_17gmr}, but for SN 2001dh.}
    \label{fig:app_01dh}
\end{figure*}

\end{document}